\documentclass[sigconf]{aamas}  
\pdfoutput=1
\usepackage{booktabs}
\usepackage{soul}
\usepackage{url}
\usepackage{subfigure}
\usepackage{algorithm}
\usepackage{algorithmic}
\usepackage{float}
\usepackage{graphicx}

\begin{document}

\title{Learning Shaping Strategies in Human-in-the-loop Interactive Reinforcement Learning}  



\author{Chao Yu}
\affiliation{%
  \institution{School of Computer Science \& Technology, Dalian University of Technology}
  \streetaddress{Dalian University of Technology}
  \city{Dalian, 116024}
  \state{China}
  \postcode{116024}
}
\email{cy496@dlut.edu.cn}

\author{Tianpei Yang}
\affiliation{%
  \institution{School of Computer Science  \& Technology, Dalian University of Technology}
  \streetaddress{Dalian University of Technology}
  \city{Dalian, 116024}
  \state{China}
  \postcode{116024}
}
\email{1464439923@qq.com}

\author{Wenxuan Zhu}
\affiliation{%
  \institution{School of Computer Science  \& Technology, Dalian University of Technology}
  \streetaddress{Dalian University of Technology}
  \city{Dalian, 116024}
  \state{China}
  \postcode{116024}
}
\email{654424653@qq.com}

\author{Dongxu Wang}
\affiliation{%
  \institution{School of Computer Science  \& Technology, Dalian University of Technology}
  \streetaddress{Dalian University of Technology}
  \city{Dalian, 116024}
  \state{China}
  \postcode{116024}
}
\email{3211423674@qq.com}

\author{Guangliang Li}
\affiliation{%
  \institution{School of Computer Science, Ocean University of China}
  \streetaddress{Ocean University of China}
  \city{Qingdao, 266100}
  \state{China}
}
\email{guangliangli@ouc.edu.cn}

\begin{abstract}  
Providing reinforcement learning agents with informationally rich human knowledge can dramatically improve various aspects of learning. Prior work has developed different kinds of shaping methods that enable agents to learn efficiently in complex environments. All these methods, however, tailor human guidance to agents in specialized shaping procedures, thus embodying various characteristics and advantages in different domains. In this paper, we investigate the interplay between different shaping methods for more robust learning performance. We propose an adaptive shaping algorithm which is capable of learning the most suitable shaping method in an on-line manner. Results in two classic domains verify its effectiveness from both simulated and real human studies, shedding some light on the role and impact of human factors in human-robot collaborative learning.
\end{abstract}

%


\maketitle

\section{Introduction}
As learning agents move from research labs to the real world, it is increasingly important for human users, even without programming skills or expert knowledge, to customize agent learning behaviors towards desired ones as quickly as possible. \emph{Interactive Reinforcement Learning} (InterRL) provides a human-in-the-loop computing paradigm that enables the integration of human knowledge (in terms of guidance, advice or demonstrations) into agent learning process such that the overall learning cost can be reduced \cite{thomaz2005real,taylor2018improving}. In InterRL, agent learning behavior is determined not only by the world state and feedback, but additionally by a real-time interaction with a human user. Understanding the impact of human factors on an InterRL process is the long-term goal of the envisioned human-agent collectives \cite{jennings2014human}.

Plenty of work has investigated how humans can help RL agents learn more efficiently through different types of interaction modes, integration methods or transferred knowledge. One typical way has been to allow humans to provide real-time feedback to the agent learning process in the form of an end-user derived reward signal \cite{taylor2018improving}. Thomaz \emph{et al.} directly combined a human reward with an RL reward in Q-learning~\cite{thomaz2006reinforcement,thomaz2008teachable}. Judah \emph{et al}.  applied an off-line labeling process to critique the agent's actions as good or bad and shape the agent's learning process using this critic \cite{judah2010reinforcement}. Knox \emph{et al.} proposed the TAMER framework to interactively shape agent behavior using a human reward function approximated by supervised learning approaches \cite{knox2008tamer}, and investigated how various combination methods influence the learning performance \cite{knox2010combining,knox2012reinforcement,knox2015framing}. Yet, recent studies showed that directly estimating the human policy can be a more effective use of human feedback  \cite{griffith2013policy,cederborg2015policy}. Other works resorted to summarization of off-line human demonstrations in shaping agent learning behaviors \cite{taylor2011integrating,brys2015reinforcement,chernova2016learning}.

Prior work develops InterRL methods that tailor the human's guidance to agents with a particular representation or underlying learning algorithm, offering effective yet specialized shaping procedures. The human knowledge, expressed in terms of either  feedback reward signals, estimated policies, or summarized decision rules from demonstrations, is integrated into agent learning process by shaping a specific component of RL, i.e., the value function, action, reward or policy \cite{abel2017agent}.  While this specific design can leverage insights about the learning algorithm or representation to derive more powerful InterRL methods, it inevitably faces generalization and interpretation issues. The methods may perform well for some types of algorithms or domains, but poorly on others, thus, general observations or conclusions claimed in previous studies may not hold consistently. For example, previous studies have showed that methods in which human rewards directly shape an RL agent's actions and policies are more efficient than others~\cite{knox2010combining}. This claim, however, may be only valid when the human is more error free and has an enough large influence on biasing the agent's learning process. In fact, even for a single combination method, different weights of human rewards can cause great sensitivity in the learning performance \cite{knox2012reinforcement}. Thus, without substantial effort and further hand-engineering, existing methods usually cannot be readily applied in various domain settings.

Since different kinds of InterRL methods embody various characteristics and advantages in different domains, it is natural to explore the combinatoric space of these learning methods in order to derive more robust InterRL methods. In fact, it was recently hypothesized by researchers that the interplay between different InterRL methods would potentially lead to new powerful shaping methods by taking advantage of the benefits of each InterRL method \cite{abel2017agent}. In this paper, by summarizing and formally defining the four different kinds of InterRL methods in the literature, we distinguish how human rewards can be explicitly represented and integrated into an agent's RL process. To grasp a better understanding of the roles and advantages of different InterRL methods, we then propose an adaptive shaping algorithm that is capable of learning the most suitable InterRL method from a portfolio of different types of InterRL methods in an adaptive manner. Results from both simulated and real human studies in two classic RL problems, i.e. Pac-Man and Cart-Pole, confirm the effectiveness of our algorithm under various conditions, compared to those approaches using only one specific InterRL method. By analyzing different ways of integrating human knowledge into an agent's RL process, our work provides valuable insights into understanding the role and impact of human factors in human-robot collaborative learning.

Section 2 gives a brief introduction to RL and summarizes the general InterRL methods. Section 3 elaborates on the details of our algorithm. Section 4 introduces experimental domains and design methodology, while Section 5 gives the results. Section 6 discusses some related work, and finally, Section 7 concludes the paper.

\section{Interactive RL Methods }\label{sec:methods}
A Reinforcement Learning (RL) problem \cite{sutton1998reinforcement} is typically modeled as a Markov Decision Process (MDP), which can be defined by the tuple $(S, A, T, R)$, where $S$ and $A$ are respectively the set of agent's states and actions, $T(s, a, s') = P(s'|s, a)$ gives the probability of jumping to the new state $s'$ given current state $s$ and action $a$, and $R(s,a)$ is the reward function. An agent's behavior is defined by a policy $\pi (s,a)$ which maps states to a probability distribution over the actions. In this paper, we assume that the policy is stochastic. The aim of RL is to find the best policy $\pi^* (s,a)$ to maximize the cumulative discount reward $G=\sum_{t} \gamma^{t} r_t$,  where $r_t$ is the reward at time $t$, and $0 < \gamma < 1 $ is a discount factor.

Given an MDP and an agent's policy, we can then define the action-value function $Q^\pi (s,a)$, which represents the cumulative discount reward after taking action $a$ in state $s$ and then using policy $\pi$ to explore. According to the Bellman equation, the optimal action-value function is given by:
\begin{equation}
	Q^* (s,a) = R(s,a) + \gamma \sum_{s'} P(s'| s,a) \max_{a'} Q^* (s',a'),
\end{equation}
 and the optimal policy $\pi^* (s,a)= \mathop{\arg\max}_{a} Q^* (s,a)$.

An RL process can be characterized by four main components: the action $A$, the policy $\pi$, the reward $R$ and the value function $V$. The existing InterRL methods convert human/agent feedback signals into a reward, a value or a decision rule, and  pass the knowledge about one specific component of RL into the agent's learning process for performance acceleration \cite{abel2017agent}. Without lose of generality, four distinct types of InterRL methods can be categorized: (1) \emph{Action-based Methods:} that use the human's feedback signals to directly affect the agent's action selection process~\cite{taylor2011integrating,abel2015goal,sherstov2005improving}; (2) \emph{Policy-based Methods}: that combine the human's policy with the agent's policy to influence the agent's decisions~\cite{fernandez2006probabilistic,griffith2013policy,cederborg2015policy}; (3) \emph{Reward-based Methods:} that convert the human's feedback signals directly into some useful form to be integrated with the agent's reward~\cite{devlin2012dynamic,brys2015reinforcement,thomaz2006reinforcement}; and (4) \emph{Value-based Methods:} that combine the human's value functions (e.g., $V(s)$ or $Q(s,a)$) with the agent's learnt value functions~\cite{taylor2011integrating,knox2010combining}.


\section{Learning Shaping Strategies}

Previous studies have shown that the above InterRL methods have distinct advantages for different RL tasks or at different learning stages~\cite{knox2012reinforcement}. It is thus necessary to investigate how different kinds of InterRL methods can interact with each other, and how this interplay can impact the final learning performance \cite{abel2017agent}. To this end, we propose an algorithm to adaptively shape an RL agent with human knowledge. Our algorithm takes several different InterRL methods as input, and maintains the cumulative rewards that they can achieve during the current learning phase. Then, it chooses the best method in the current stage to execute. In order to enable exploration, the cumulative reward of each method is converted into a probability, and \emph{importance sampling} is applied to speed up the update of probability in choosing each method.



\begin{algorithm}[htb]
	\caption{Learning Shaping Strategies in InterRL}
	\begin{algorithmic}[1]
		\REQUIRE $M;$
		\STATE $w_{m_i} \leftarrow 0, \quad i = 1, 2, \dots, |M|;$
			\WHILE {not converge}
		\STATE $P(m_i ) = \frac{e^{\beta (w_{m_i}-min(w_{m_i}))} }{\sum_{j=1}^{|M|} e^{\beta (w_{m_j}-min(w_{m_i}))}};$ \\
		\STATE sample a method $m_{current}$ from $P(m_i ), \quad i = 1,2,\dots,|M|;$
		\STATE $sim_{m_{i}} \leftarrow 1, \quad i = 1, 2, \dots, |M|;$
		\STATE $R \leftarrow 0;$
		\STATE $s$ = \textbf{init\_task}();
		\WHILE {episode not over}
		\STATE get human feedback $H$;
		\STATE $a$ = \textbf{get\_action}($m_{current},s,H$);
		\STATE $s, R_e$ = \textbf{do\_action}($s$, $a$);
		\STATE \textbf{update\_q}($s, a, R_e, H$);
		\STATE $R \leftarrow R + R_e$;
		\FOR {$i = 0$ to $|M|$}
		\STATE $\pi_{m_i }$ = \textbf{get\_prob}$(m_i, s, a, H)$;
		\STATE $sim_{m_i } = sim_{m_i }* \pi_{m_i }$;
		\ENDFOR
		\ENDWHILE
		\FOR {$i = 0$ to $|M|$}
		\STATE $w_{m_i } = w_{m_i } + \tau *\frac{sim_{m_i }} {\sum_{j=1}^{|M|} sim_{m_j }} (R-w_{m_i })$;
		\ENDFOR
		\ENDWHILE
	\end{algorithmic}
\end{algorithm}

Assuming a set of InterRL methods ($M={m_1,\dots, m_{|M|}}$), \textbf{Algorithm 1} gives the main procedure of our algorithm, in which $w_{m_i}$ stands for the weight of method $m_i$, $sim_{m_i}$ stands for the similarity value of method $m_i$, $R$ is the accumulated reward of current running method, $R_e$ is the reward from the environment and $H$ is the reinforcement signal from the human. Initially, we set all $w_{m_i}= 0$ to indicate an equal probability of choosing each InterRL method (Line 1).
Then, the algorithm runs the chosen method and updates the weights of all the methods based on the reward obtained by that method. More specifically, in each iteration, the algorithm calculates the probability of each InterRL method using a softmax function according to the current weight vector (Line 3), and then selects a method $m_{current}$ according to the probability (Line 4). Then, the chosen InterRL method runs for one episode and returns the similarity $sim_{m_i}$ for each method $m_i$ and total accumulated reward $R$ for the current running method $m_{current}$ (Line 9-17).

At each time step in an episode, the specific InterRL method asks for the guidance from human (Line 9), i.e. the human reinforcement signal $H$ (it is possible that the human may not provide guidance at this request), and then determines the next action $a$ based on the currently chosen InterRL method $m_{current}$, the current state $s$ and human guidance $H$ (Line 10). After taking this action, the next state $s$ and reward from environment $R_e$ are received (Line 11). Then the Q-function is updated using the chosen InterRL method (Line 12), and the accumulated reward $R$ is added by the RL reward $R_e$ (Line 13). At each time step, the algorithm calculates the similarity between $m_{current}$ and other InterRL methods, based on the probability that the current action is considered to be optimal by each  InterRL method (Line 14-17). Function \textbf{get\_prob}$(m_i, s, a, H)$ outputs a probability of choosing action $a$ in state $s$ using another  InterRL method $m_i$. Since some methods such as \emph{policy-based} and \emph{action-based} methods can also influence the final policy, the human feedback $H$ should also be included in the \textbf{get\_prob} function. When the whole episode finishes, the algorithm returns the $sim_{m_i}$ value of each InterRL method and the total environmental return $R$. Since similarity $sim$ maintains the policy similarity between each method and the current running method, the weights of all InterRL methods can be updated based on $sim$ and $R$, using a TD-like learning update procedure with learning rate $\tau$ (Line 20).

\begin{figure}[h]
\centering
\includegraphics[width=0.5\textwidth]{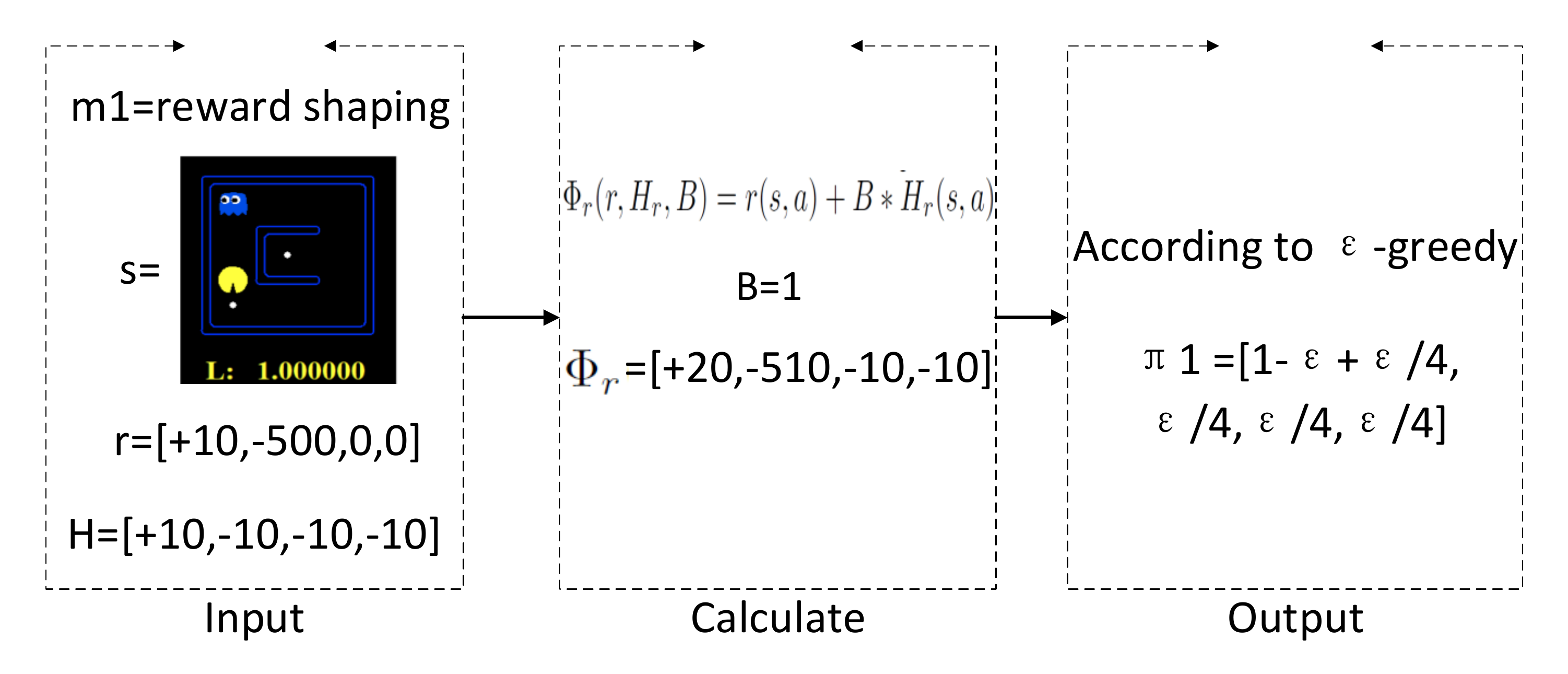}
\centering
\caption{Illustration of how to implement the get\_prob() function using reward shaping method.}
\label{fig:getprob}
\end{figure}

Figure \ref{fig:getprob} gives an illustration of how to implement the get\_prob() function using reward shaping method, where $r$ indicates the agent's reward regarding four actions (i.e., +10, -500, 0, and 0 for moving \emph{up}, \emph{down}, \emph{left} and \emph{right}, respectively), while $H$ indicates the human's reward signal regarding the four corresponding actions  (more details regarding the human reward signal $H$ will be given in Sec. 4.2.1). Assuming the  shaping parameter $B=1$, the final reward of the agent $\Phi_r (r, H_r,B)=r(s,a) + B*H_r (s,a)$ then can be computed as [+20,-510,-10,-10]. Based on this reward vector and the $\epsilon$-exploration strategy, the probability get\_prob($m_i,s,a,H$) that another method $m_i$ chooses the action $a$ adopted by the current method $m_{current}$ can be computed. As in Figure \ref{fig:getprob}, if current method $m_{current}$ has adopted action moving \emph{down}, get\_prob($m_1,s,down,H$) is then $\epsilon/4$ because the best action suggested by $\Phi_r $ is moving \emph{up}.

\section{Experiment}
We evaluate the adaptive shaping algorithm using both simulated environments and real user studies in two benchmark RL tasks that have been widely adopted in previous research. Similar to prior work \cite{griffith2013policy,mandel2017add,amir2016interactive}, the simulated environments allow us to compare methods in an inexpensive yet well-controlled setting, covering a wide range of parameter settings, while the real user studies enable us to investigate what really happens when methods are driven by humans with realistic constraints and uncertainties.

\begin{figure}[t]
\centering\includegraphics[width=0.45\textwidth]{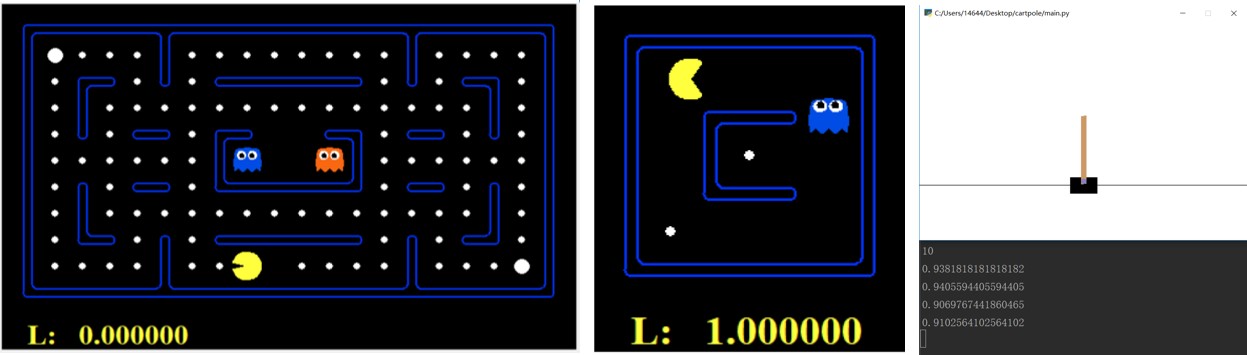}
\caption{Benchmark domains. The left two are respectively the medium Classic and small Grid version of Pac-Man. The right is the Cart-Pole domain.}\label{fig:domain}
\end{figure}

\subsection{Domain}
The two benchmark RL domains are Pac-Man and Cart-Pole as shown in Figure~\ref{fig:domain}. Pac-Man is a 2D grid game, which includes food, walls, ghosts and a Pac-Man agent. The goal of the game is to control the agent to get all the food while avoiding colliding with any ghost. Eating all food pellets ends the game with +500 reward, and being killed by the ghost ends the game with -500 reward. Each food pellet gives +10 reward, and each time step of the game costs Pac-Man -1 penalty. The actions of the agent include four movements: up, down, left, and right. When the agent encounters a wall, it stops moving and stays where it was. The state of the agent includes the agent's current position, the ghost's current position, the ghost's current direction and the presence of foods. We use the Pac-Man project developed by UC Berkeley\footnote{http://ai.berkeley.edu/project\_overview.html} and adapt it to meet requirements of our evaluation. Although different sizes of Pac-Man game are available, from medium Classic to small Grid, we base our evaluation on the small Grid domain (5*5 square), which contains two foods and a ghost (the same setting as in \cite{griffith2013policy}). The medium Classic size is prohibitively large for real human studies. In fact, we test that the average time for one run in a medium Classic domain is nearly 11.3 times longer than that in a small Grid domain. Considering that even one run in the small Grid domain takes averagely 1.5 hours, direct evaluation in the medium Classic domain by real humans is thus impracticable.

The Cart-Pole is applied to test the performance of methods in continuous domains. In this domain (right part in Figure~\ref{fig:domain}), a small cart can move around with a pole standing on the cart. The agent needs to choose two movements: moving left or right to control the cart, while keeping the pole upright. The agent gains +1 reward by maintaining the pole at each time it interacts with the environment.  When the pole is unable to remain vertical, the game ends with a -1 reward. An episode ends if the pole stands upright for 200 steps. The continuous state of the agent includes the angle and the acceleration of the pole. Interactive learning in such continuous settings is a challenging task, especially when dealing with human factors such as delayed reaction time, limited patience, and cognitive cost etc. \cite{vien2012reinforcement}.

\subsection{Design}

\subsubsection{Explicit InterRL Methods}
There are several ways of deriving a specific InterRL method as described in Section~\ref{sec:methods}. Following previous studies \cite{knox2010combining,knox2012reinforcement}, four explicit methods are chosen in our evaluation: the \emph{Action Biasing}, the \emph{Control Sharing}, the \emph{Reward Shaping} and the \emph{Q Augmentation}. Let $H_a, H_\pi , H_r, H_v$ denote the human's shaping function regarding actions, policies, rewards, and value functions, respectively, and $B$ denote a predefined shaping parameter. The four InterRL methods can be formally given as follows:

\begin{description}
  \item[\emph{Action Biasing (AB):}] AB is a typical action-based method, in which the action chosen by the agent is determined by $\Phi_a (c, H_a, B) = argmax_a [Q(s,a)+B*H_a (s,a)]$, where $H_a (s,a)$ can be the human's value function on actions.
  \item[\emph{Control Sharing (CS):}] Similar to the \emph{Probabilistic Policy Reuse} (PPR) strategy \cite{fernandez2006probabilistic}, CS is one typical policy-based method. Instead of providing an advice directly to the agent, CS controls the agent' policy of choosing actions by $\Phi_\pi (\pi,H_\pi,B)=\pi(s,a)^l * H_\pi (s,a) ^{1-l}$, where $l$ is a Bernoulli random variable that controls the probability of choosing human's policy function.
  \item[\emph{Reward Shaping (RS):} ] RS is a classic reward-based method, in which the reward of agent can be determined by $\Phi_r (r, H_r,B)=r(s,a) + B*H_r (s,a)$.
  \item[\emph{Q Augmentation (QA):}]  QA can be considered as one type of value-based methods, in which the Q value of agent can be given by $\Phi_v (v, H_v, B)=Q(s,a)+B*H_v (s,a)$.
\end{description}

\begin{figure}[t]
\centering
\includegraphics[width=0.4\textwidth]{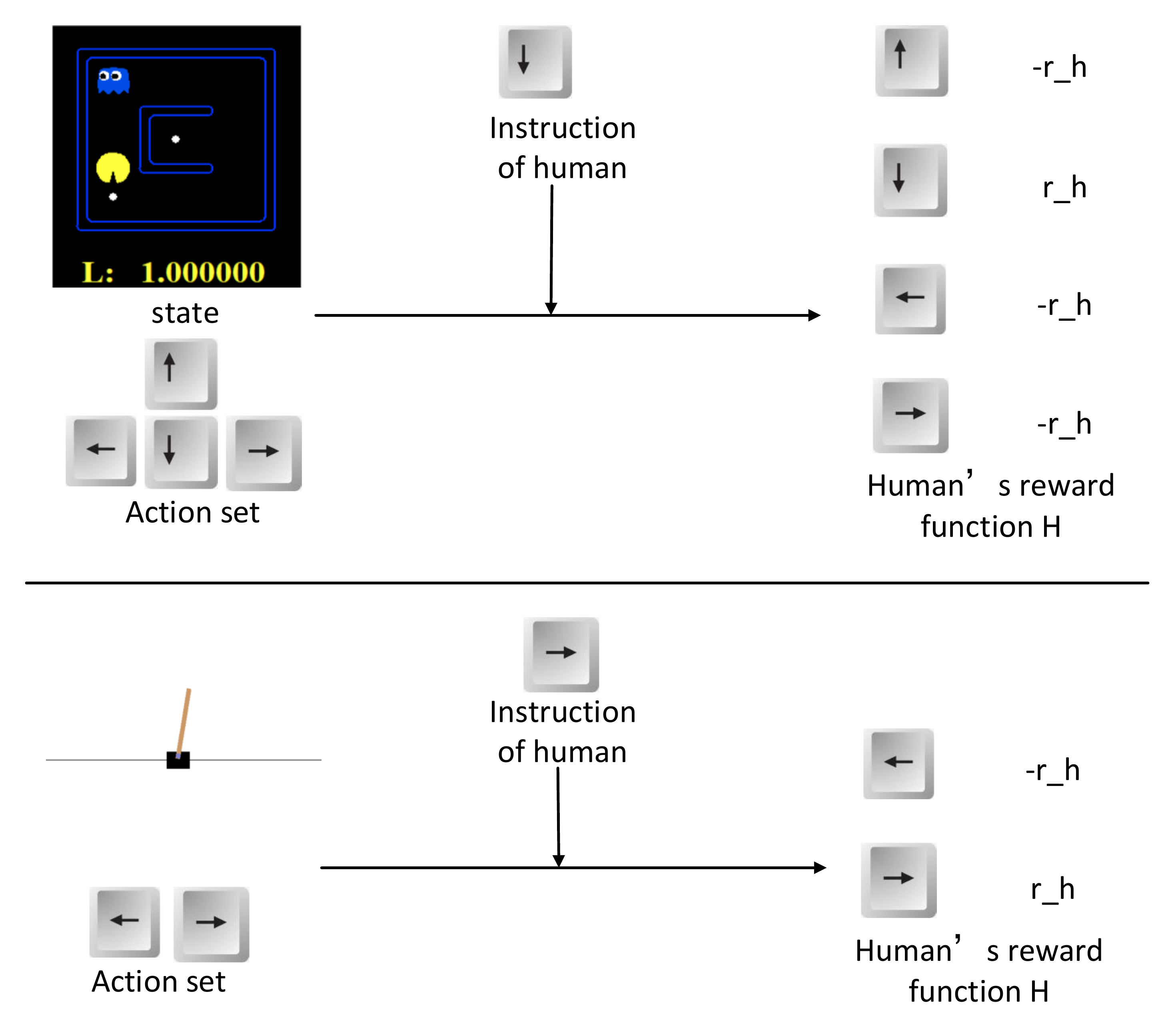}
\centering
\caption{Illustration of the human reward function $H$ in the Pac-Man and Cart-Pole domain.}
\label{fig:H}
\end{figure}

A human has only two instructions: right or wrong.
A reward vector is used to denote the human's reward function $H$ ($H_a, H_\pi , H_r, H_v$) regarding the action choices. When the human thinks that an action is the best choice, the corresponding action in the reward vector is denoted as $r_h$, and other actions are denoted as $-r_h$. Figure \ref{fig:H} illustrates the formulation of human reward function $H$ in the Pac-Man and Cart-Pole domain, where only the action consistent with the one suggested by the human is set to a positive value $r_h$, while other actions are set to a negative value $-r_h$. Therefore, the value of $r_h$ indicates the magnitude of human influence on the agent's learning process. In Pac-Man, the discount factor $\gamma=0.7$, updated parameter $\alpha=0.3$ and exploration parameter $\epsilon=0.1, \beta = 5$. The shaping parameter in all four methods $B=1$ and this value decreases by $\frac{1}{30000}$ after each episode. Each method performs 30,000 episodes (this is because ordinary Q-learning needs roughly 30,000 episodes to converge), and final results are averaged over 20 independent runs. In Cart-Pole, discount coefficient $\gamma = 0.99$, and exploration parameter $\epsilon = 0.3, \beta = 5$. The parameter $B = 1$ in all methods, and decreases by $\frac{1}{2000}$ after each episode. Each of the methods performed 2000 episodes, and final results are averaged over 20 independent runs.

\subsubsection{Simulated Oracles}
Following~\cite{griffith2013policy}, we obtained the optimal policy using standard Q-learning in the two experiments to simulate completely correct human guidance. When the agent requires human guidance, we then use the largest Q-value as the perfect human guidance in current state. It should be noted that, although the Q-learning is used to obtain the optimal Q values to model human guidance, in the experiments, only \emph{value-based} methods can directly access these Q values. Other methods can only receive the information of optimal actions associated with the maximum Q values by $a^* = argmax_a Q[s,a]$. In order to quantitatively study the advantages and disadvantages of different methods in various parameter settings, two extra parameters are introduced as in~\cite{griffith2013policy,cederborg2015policy}: $L$ (the \emph{likelihood} of feedback) which represents the probability that the human provides guidance at each time step, and $C$ (the \emph{consistency} of feedback) which represents the probability that the human provides the optimal instructions correctly. The benefit of simulated oracles is that its allows us to compare methods in an inexpensive yet well-controlled setting, covering a wide range of parameter settings to model human errors and limited patience through the two parameters of consistency $C$ and likelihood $L$.


\subsubsection{Real Humans}
Compared to learning autonomously, learning efficiently from human inputs brings significant challenges due to issues such as humans' reaction delay, inconsistent or ambiguous inputs, and limited patience and attention \cite{peng2016need,thomaz2006reinforcement}. Despite these inherent challenges, there are many more practical issues that arise when implementing an InterRL system with real humans. For example, how to design a user interface to enable seamless interactions between agents and humans? What kind of visualizations causes humans to develop the most useful actions in response to a request? How to reconcile the speed gap between autonomous learning and human learning? When and how frequently should humans provide guidance given their limited patience and short period of attention? As the \emph{consistency} of feedback $C$  by itself is embodied in the errors during human learning process, we pay special attention to the modelling of \emph{likelihood} of feedback $L$ in human user studies. Since $L$ is a probability that cannot be explicitly modeled during human interactions, we adapt the original game interface by clearly visualizing the parameter of $L$ (see Figure~\ref{fig:domain}) to indicate the percentage of steps in which humans have provided inputs so far. To this end, we first let the participants play the games several times before formally starting the experiment, and estimate the total steps required to finish one run of play. In the studies, $L$ is set to 0.01 by default to reduce the amount of human focus on the whole training process.

For human teachers, constantly paying attention imposes cognitive costs and can be simply unrealistic. Even if the teacher is a computer agent, transmitting the agent's every state to the teacher can cause a prohibitive communication cost. We then propose three \emph{training strategies} that capture different types of human behaviors: (1) \textbf{\emph{Early-advice}}: humans provide inputs primarily in the early stage of learning; (2) \textbf{\emph{Sporadic-advice}}: humans provide inputs in a regular frequency throughout the learning period; and (3) \textbf{\emph{Late-advice}}: humans provide inputs primarily in the late stage of learning. It is noted that many recent works focus on investigating when and how engagement of a teacher (either a human or a teaching agent) is more valuable in InterRL settings, by using the evaluation of the learner's uncertainty or performance \cite{li2013using,torrey2013teaching}, or estimation of current policy of either interaction counterpart \cite{macglashan2017interactive,loftin2014strategy}. The focus of our study is simply on the performance comparison of different InterRL methods using one specific training strategy.

To solve the issues of delay in human reaction and speed discrepancy between computers and humans, the game environment can be switched into a \emph{human interaction mode} by pressing a key to slow down the refreshing frequency to humans' capability range. This provides humans with enough time to give correct feedbacks. Moreover, as the human feedbacks are always consistent during a certain number of successive steps (e.g., a constant right force should be provided if the pole is falling rightwards), the game environment is set to a \emph{default human interaction mode} when the computer takes over the human to provide the same feedback as before. This liberates humans from the long term laborious focus on the game playing. In fact, it is unrealistic for humans to play the continuous Cart-Pole by giving feedbacks in each of the 100 frames during each episode.

\begin{figure*}[tb]
	\centering
	\subfigure[$L=1, C=0.55$] {
		\includegraphics[width=0.25\textwidth]{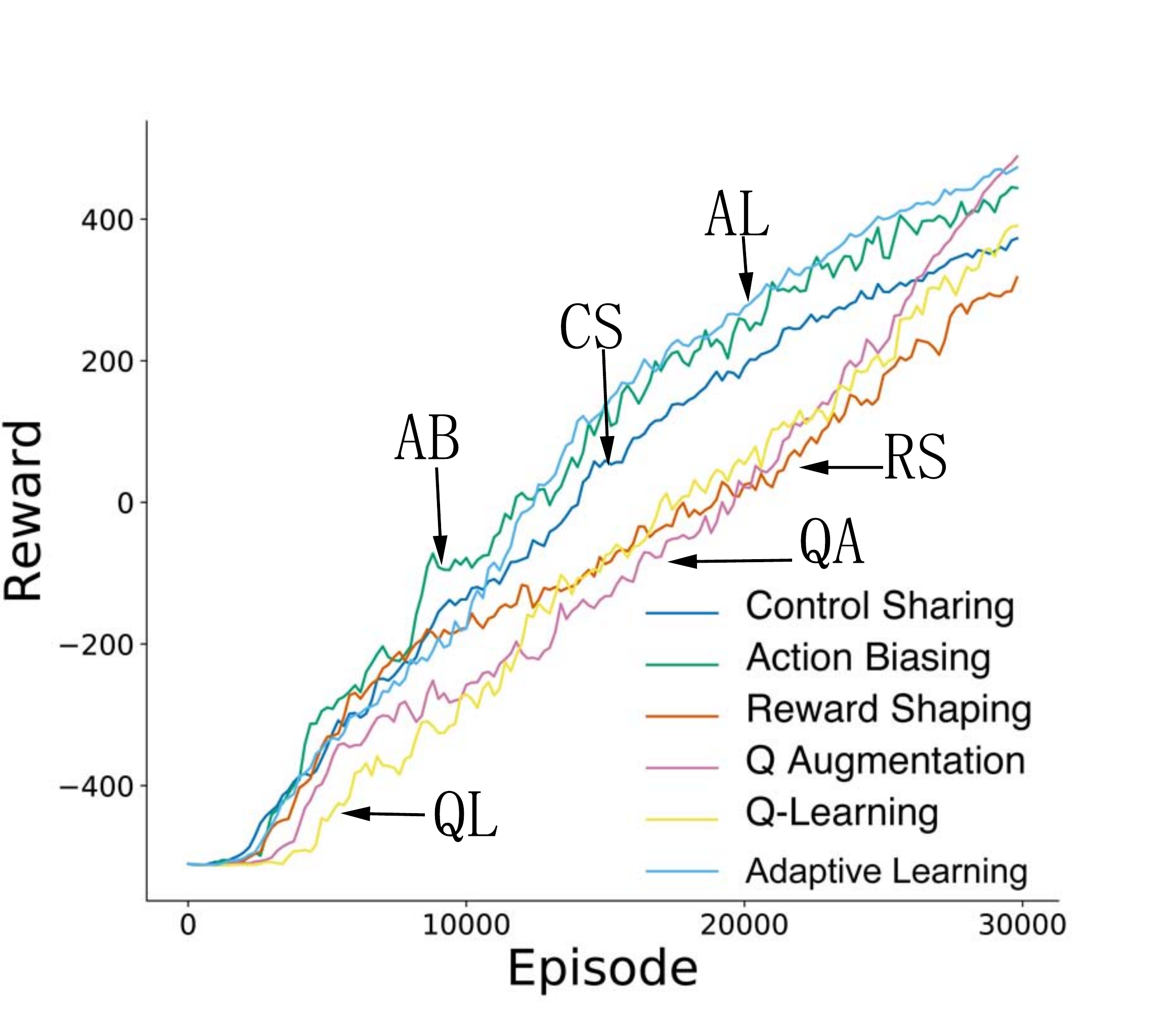}
		\label{pacman:a}
	}
	\subfigure[$L=0.01, C=0.8$] {
		\includegraphics[width=0.25\textwidth]{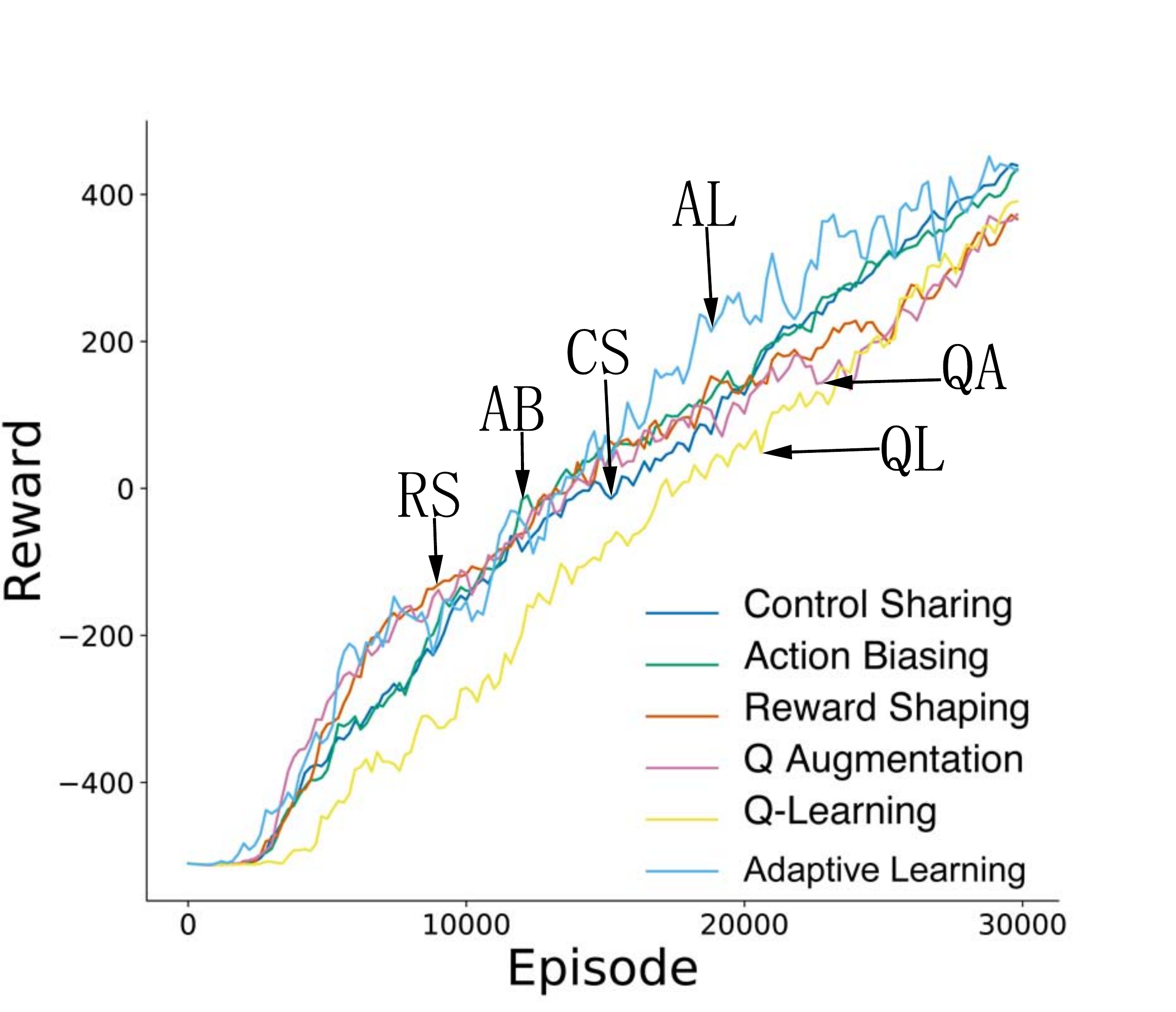}
	\label{pacman:b}
	}
	\subfigure[$L=0.1, C=0.8 $] {
		\includegraphics[width=0.25\textwidth]{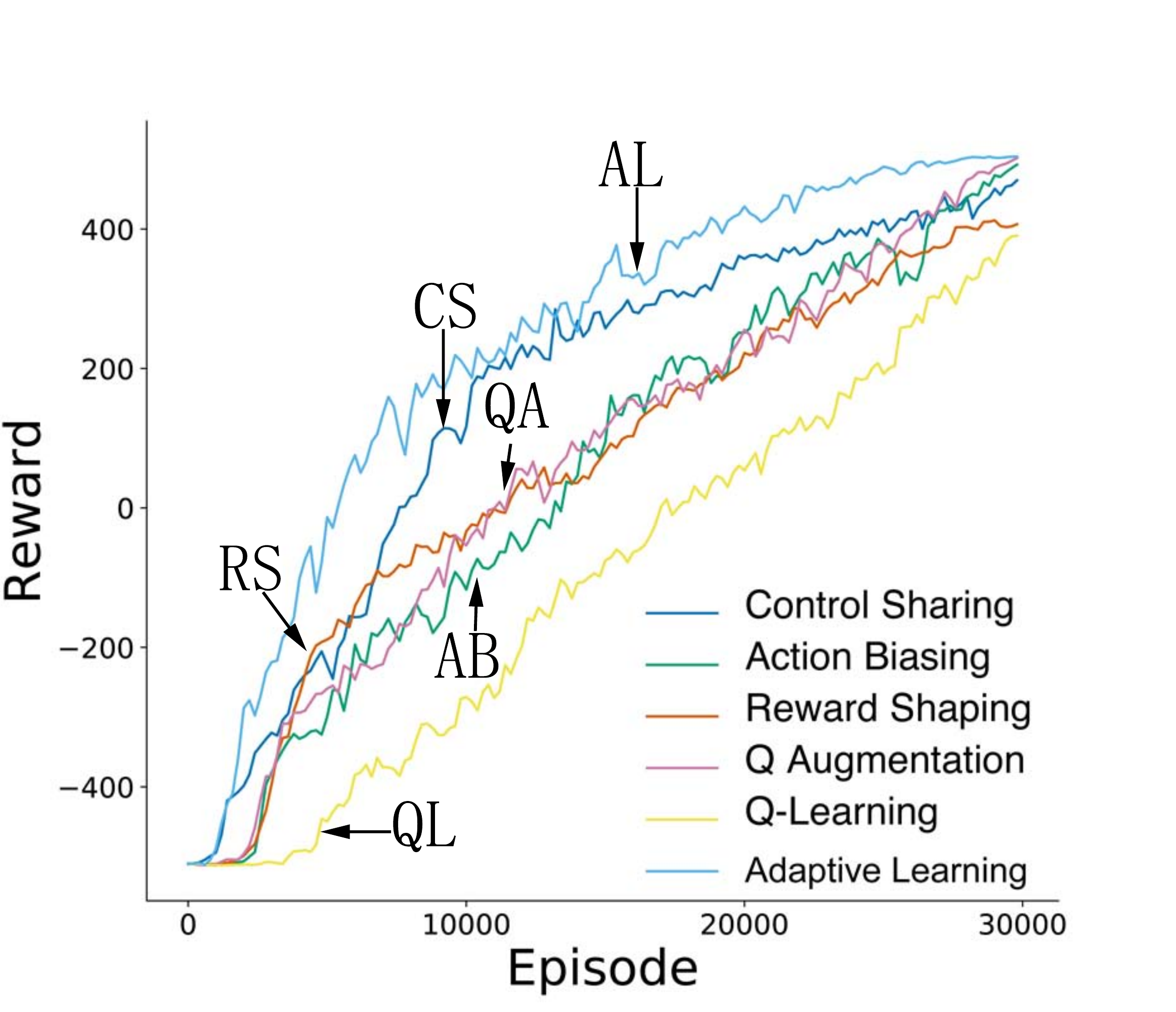}
		\label{pacman:c}
	}
	\caption{Average reward under different parameters $L$ and $C$ in the Pac-Man domain when $r_h=10$ (AL, CS, AB, RS and QA denote the \emph{adaptive learning}, \emph{control-sharing}, \emph{action-biasing}, \emph{reward-shaping} and \emph{Q-augmentation} method, respectively.).}
	\label{pacman}
\end{figure*}

\begin{figure*}[tb]
	\centering
	\subfigure[$L=0.01, C=0.8$] {
		\includegraphics[width=0.23\textwidth]{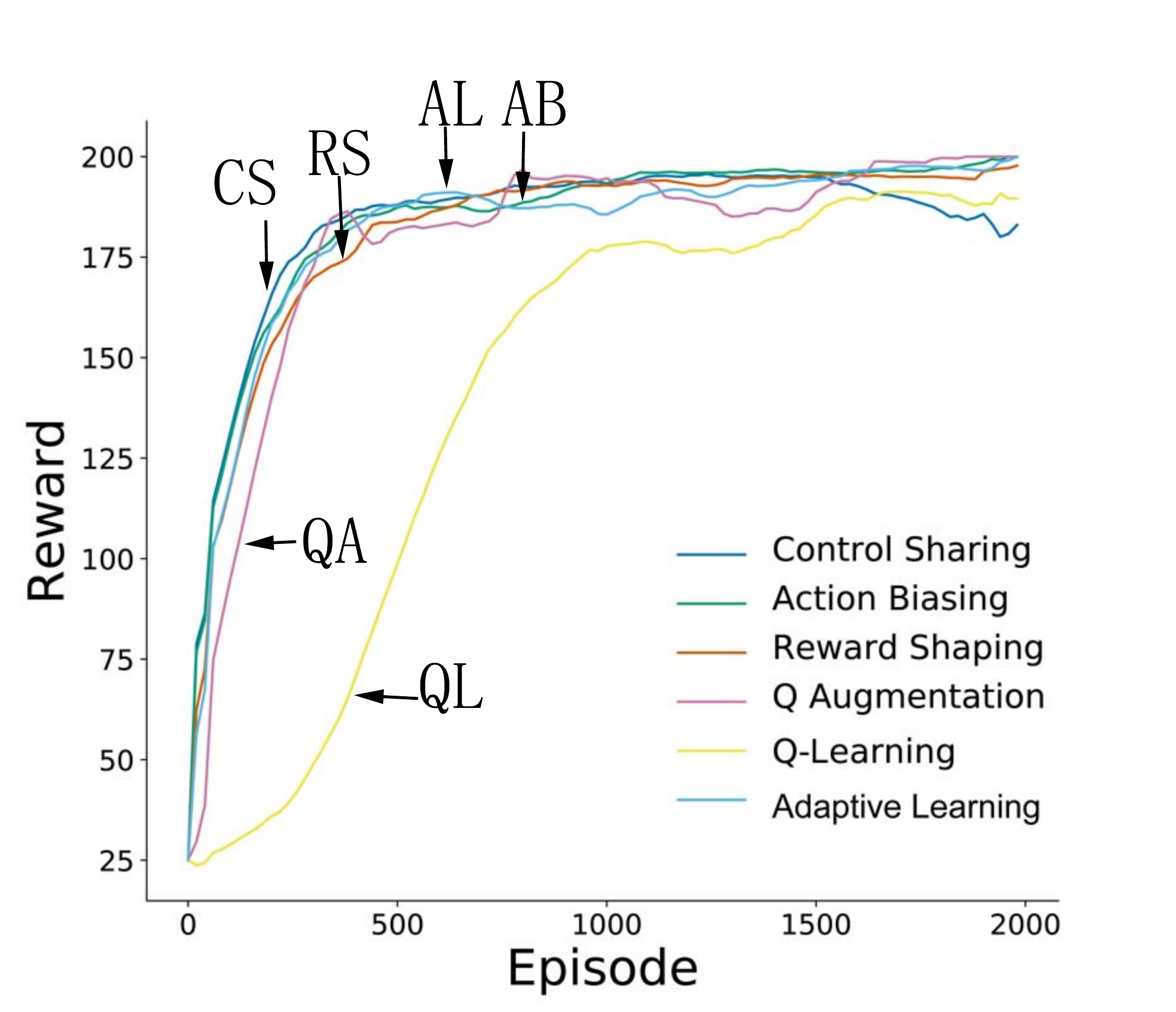}
		\label{cartpole:a}
	}
	\subfigure[$L=1, C=0.5$] {
		\includegraphics[width=0.23\textwidth]{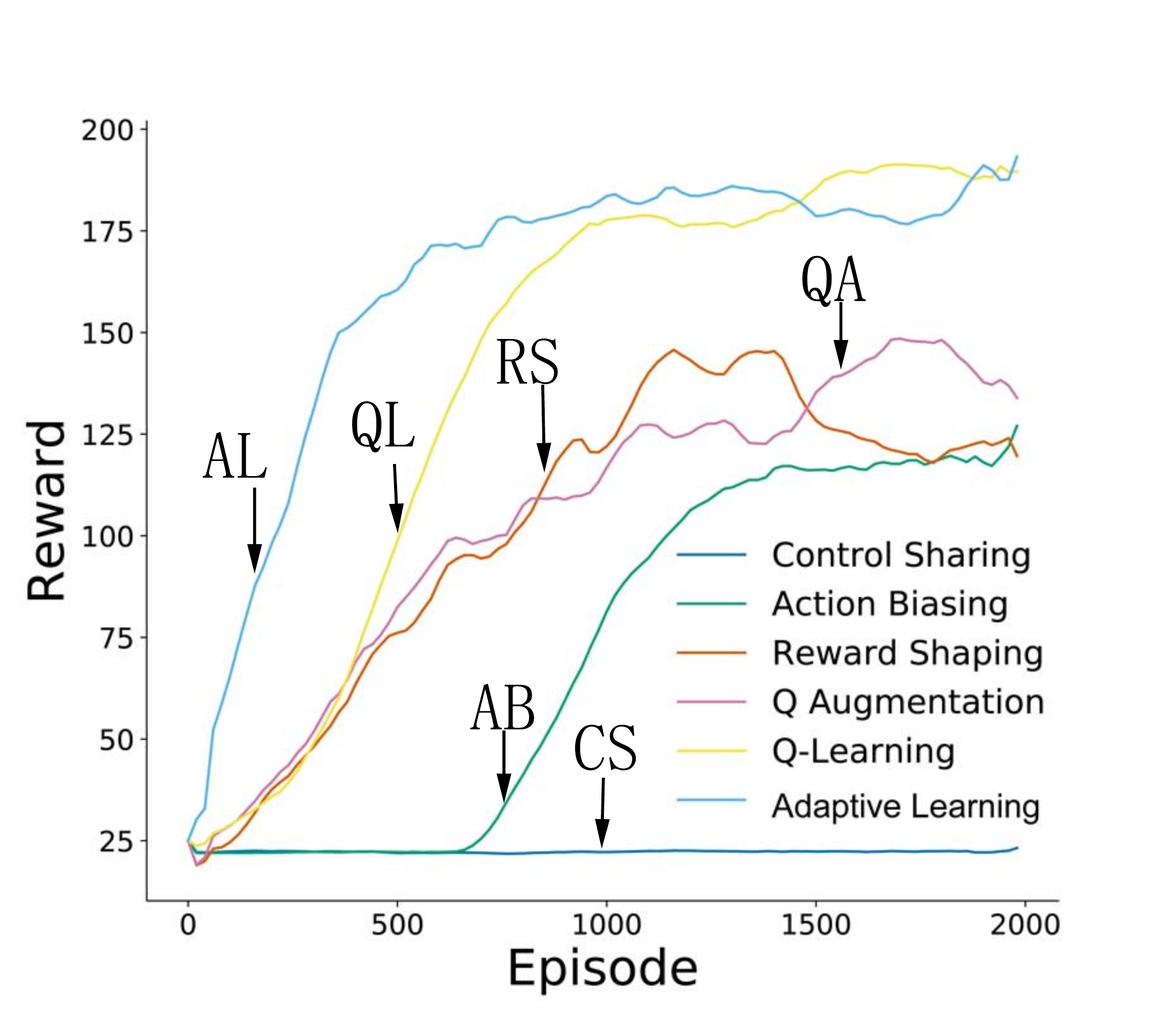}
		\label{cartpole:b}
	}
	\subfigure[$L=1, C=0.51$] {
		\includegraphics[width=0.23\textwidth]{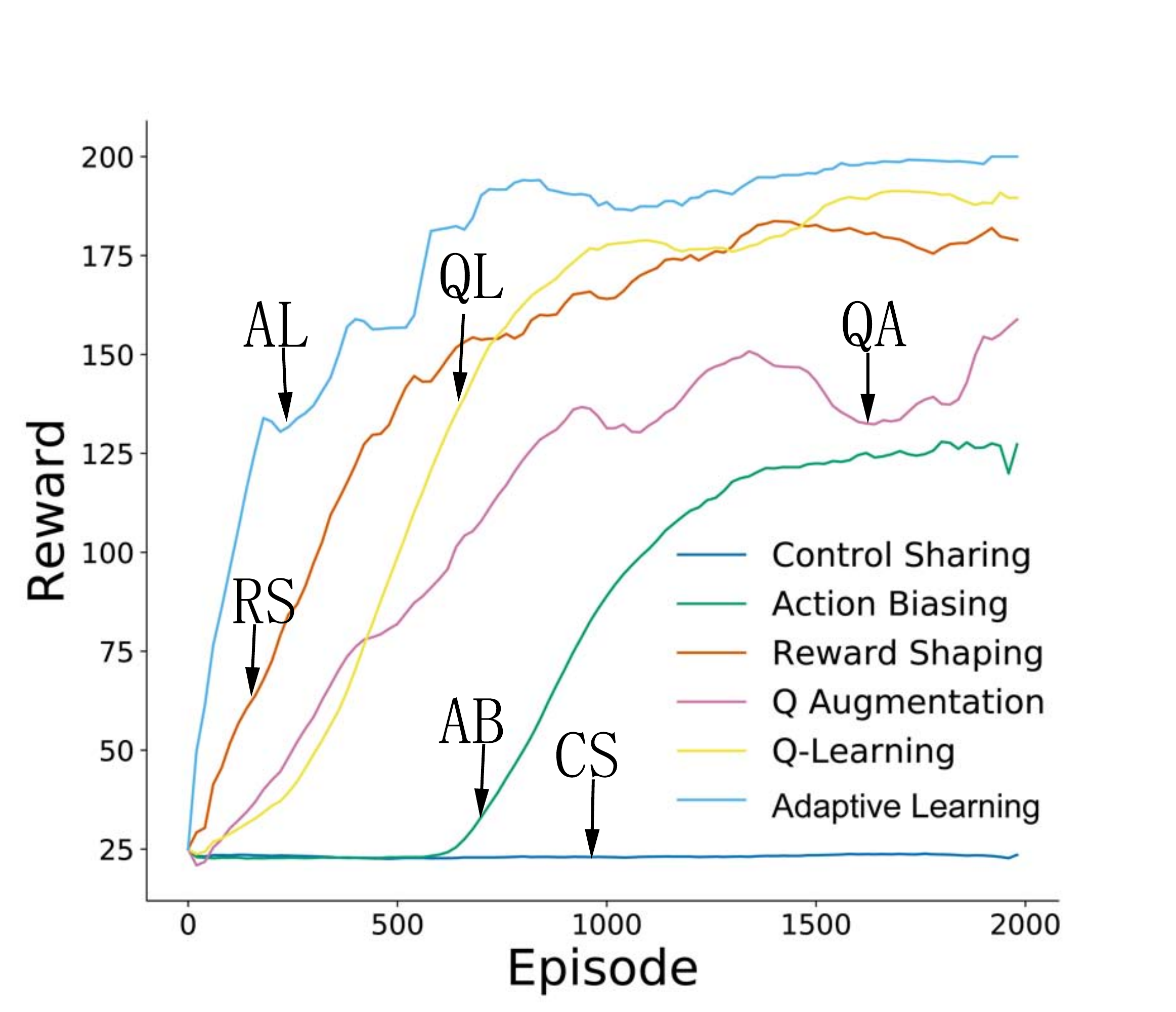}
		\label{cartpole:c}
	}
	\subfigure[$L=1, C=0.8 (r_h=100)$] {
		\includegraphics[width=0.23\textwidth]{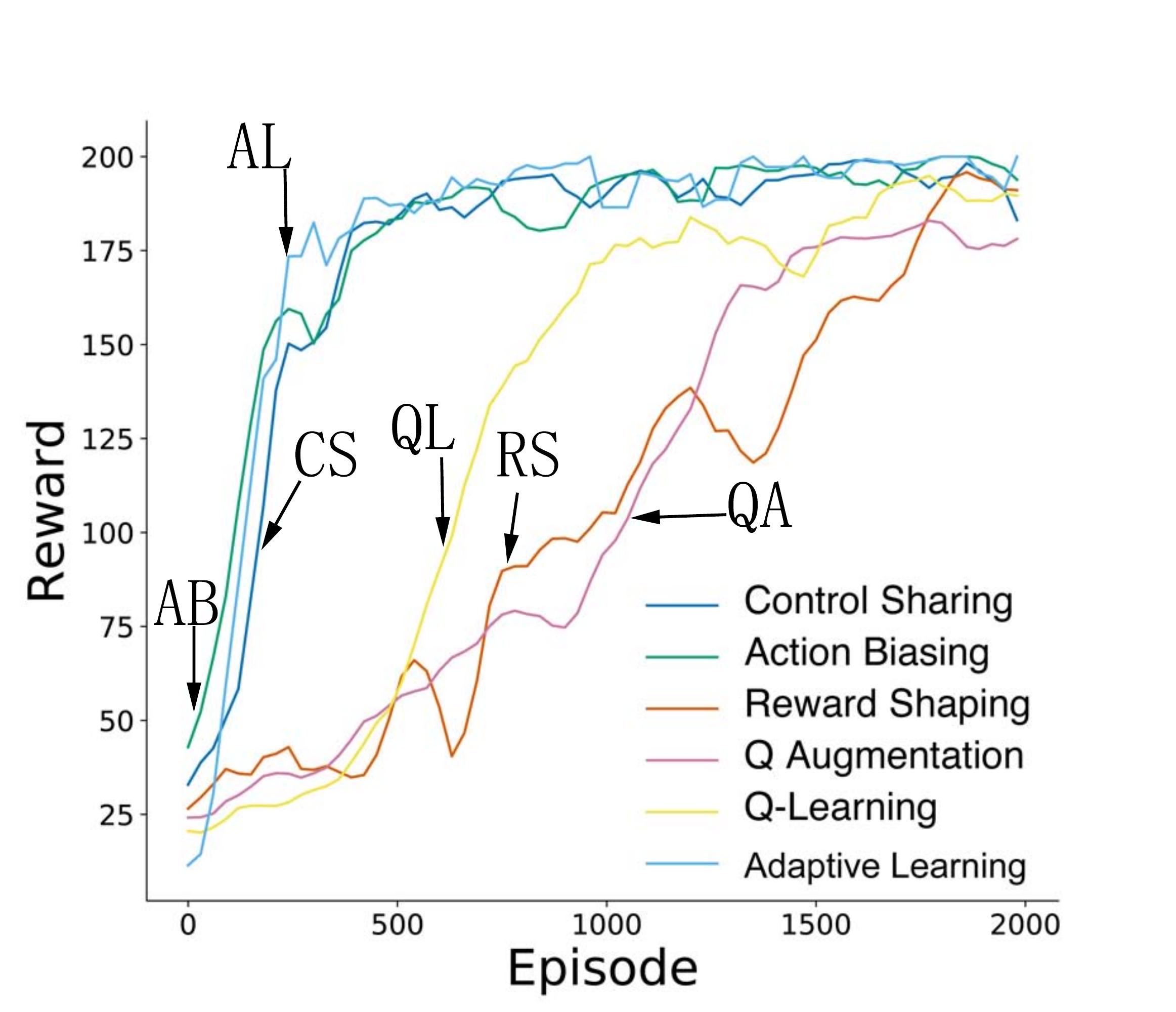}
		\label{cartpole:d}
	}
	\caption{Average reward under different parameters of $L, C$ in the Cart-Pole domain, when $r_h=10$ (a)-(c) and $r_h=100$ (d).}
	\label{cartpole}
\end{figure*}

We solicited 10 volunteers to provide data for this experiment. Each participant played the game several times to get familiar with the interface and how to interact with the RL agent first, and then formally played the game for 5 times. The final results are averaged over all the participants (i.e., 50 independent runs). Note that it normally takes 1 to 1.5 hour to finish one run of training (30000 episodes in Pac-Man and 2000 episodes in the Cart-Pole). Thus, each learning curve is the result of a long period of game playing by the participants. Unless specified otherwise, \emph{early-advice} strategy is adopted due to its better performance in terms of cost and efficiency.

\section{Results}
\subsection{Simulated Oracles}
Figure~\ref{pacman} plots the average cumulative rewards under different domain settings in the Pac-Man domain. When the likelihood of feedback $L=1$ and the consistency of feedback $C=0.55$ (a), our algorithm AL behaves slightly better than CS and AB method, while QA and RS can only achieve similar performance with Q-learning. The performance gap between all these methods is still very minor when $L=0.01$ and $C=0.8$ (b). However, when increasing human correctness to $C=0.8$ and at the same time providing human inputs only at a median frequency of $L=0.1$ (c), AL converges completely faster than the other methods.

The performance distinction of the methods is more apparent in the Cart-Pole domain as shown in Figure~\ref{cartpole}. When $L$ is 0.01 and $C$ is 0.8 (a), all the methods can achieve similar performance, enabling faster convergence than the basic Q-learning algorithm. This implies that if the human can give correct guidance with high probability ($C=0.8$), individual InterRL methods can greatly increase the learning performance, even this guidance occurs only occasionally ($L=0.01$). When $L$ is 1 and $C$ is 0.5 (b), which means that the optimal and non-optimal human actions are given randomly at each time step, the four combination methods all perform poorly. Especially, CS cannot converge at all, while AB, RS and QA all perform far worse than the traditional Q-learning algorithm. In order to further reveal the impact of human's errors on learning performance, we slightly increase the correctness probability $C$ to 0.51 (c). The result shows that a very minor increase in $C$ leads to apparent impact on the learning performance, especially for the reward shaping method, which now achieves slightly better performance against Q-learning. This result demonstrates that, to enable good performance of InterRL methods, the human must have a good understanding of the RL task, and thus can give guidance correctly (i.e., an adequately high $C$).

It is generally believed that methods shaping the agent's actions and policies (e.g., AB and CS) are more efficient than methods shaping the agent's rewards and value functions (e.g., RS and QA) \cite{knox2012reinforcement}. However, when human actions are more prone to errors, AB and CS perform far worse than RS and QA. This is easy to understand since AB and CS are methods that are directly shaped by human actions or policies, which are now error-prone. Similarly, different formulations of value functions would possibly have significant impacts on methods like RS and QA. Figure~\ref{cartpole:d} shows the result when increasing the human reward $r_h$ to 100 in order to raise the influence of humans on biasing the learning process. Although only a few human errors occur, the higher value of reward can cause more dramatic change in the reward or value functions, which in turn impairs the learning performance of RS and QA a lot.

\begin{figure}[t]
	\centering
	\subfigure[Interplay of methods] {
		\includegraphics[width=0.22\textwidth]{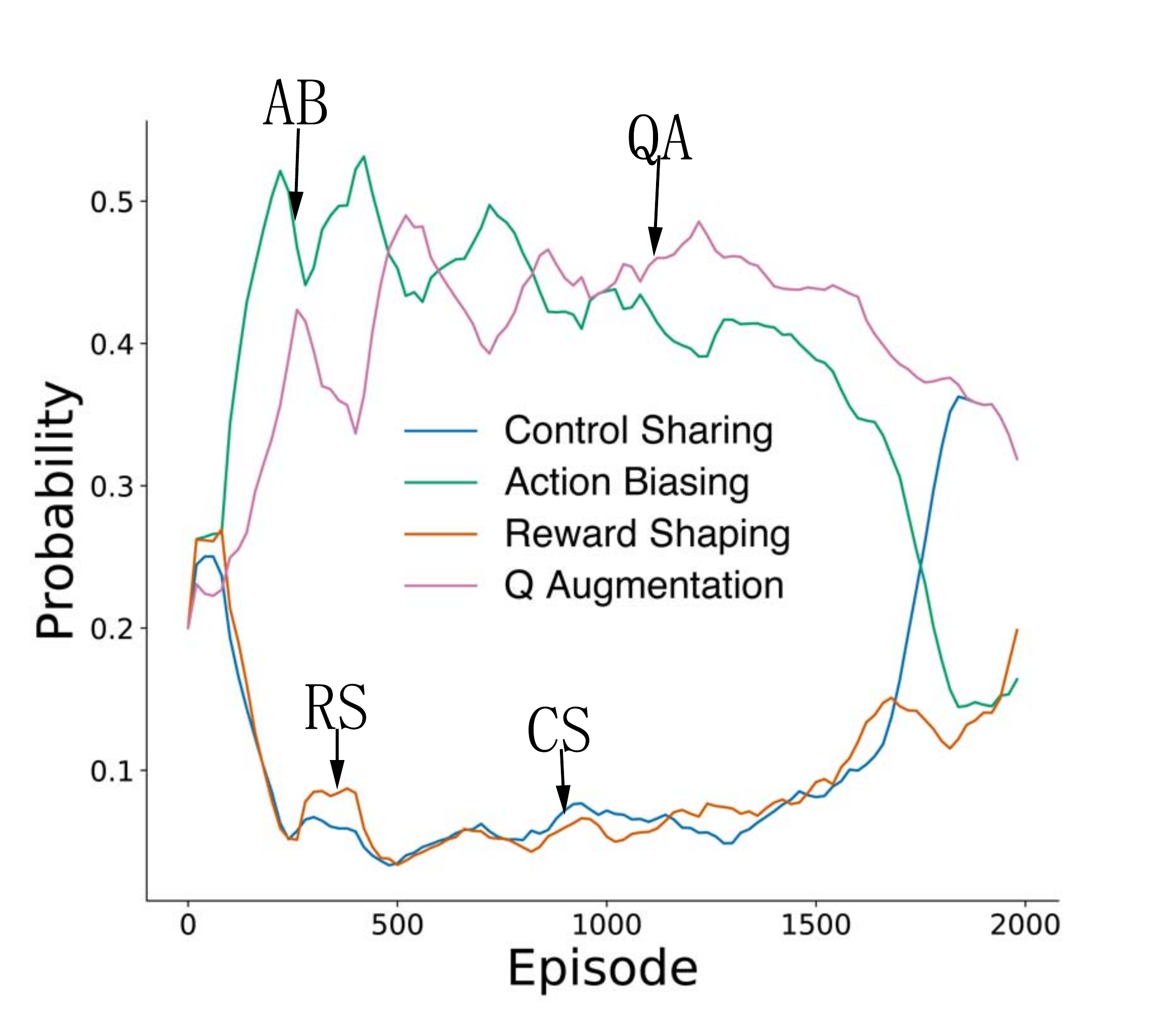}
 		\label{choose_prob}
	}
	\subfigure[In dynamic environments] {
		\includegraphics[width=0.22\textwidth]{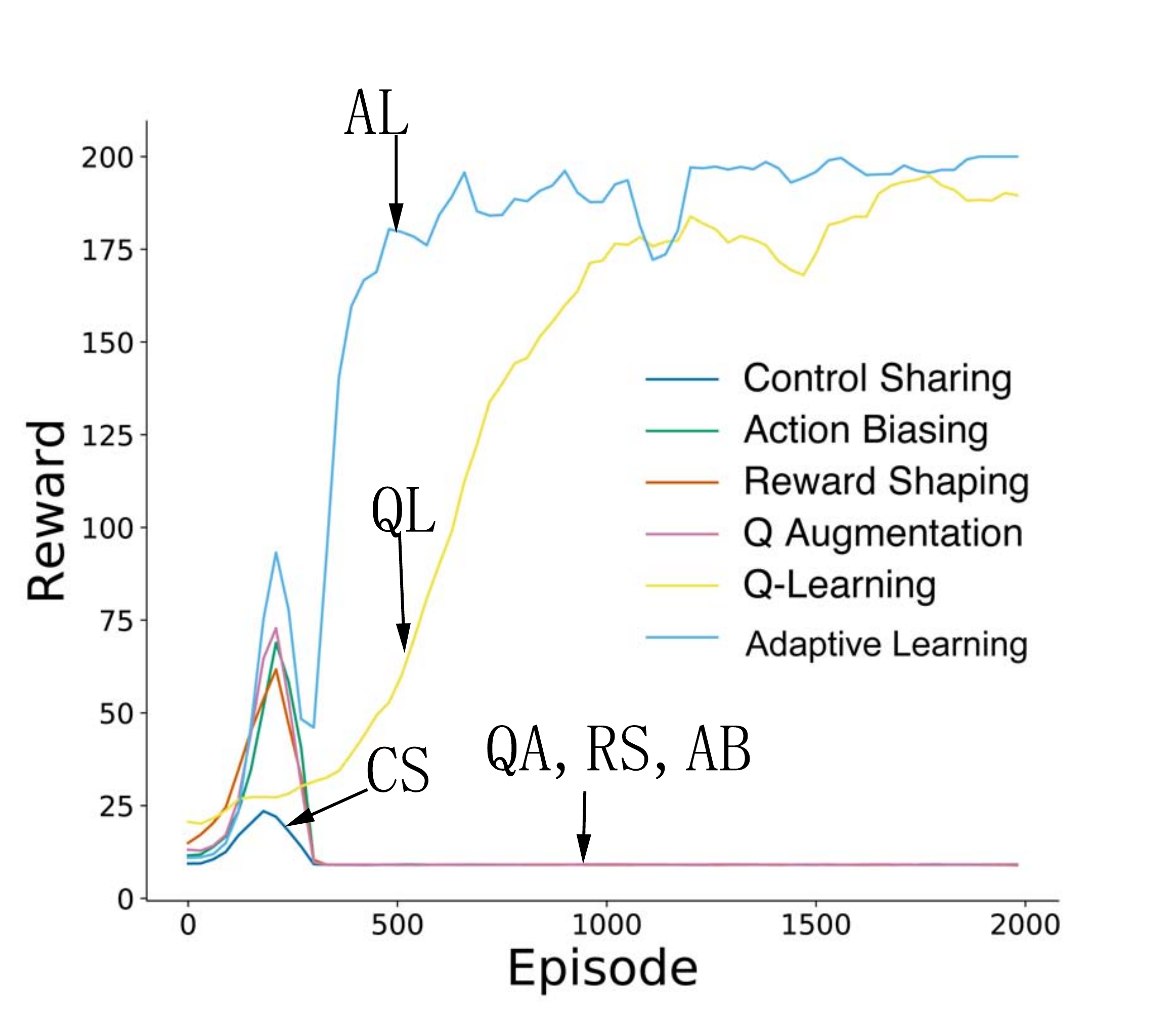}
 		\label{worst:a}
	}
	\caption{The interplay between different InterRL methods using the adaptive shaping algorithm (a) and the performance of our algorithm in dynamic environments (b).}
	\label{worst}
\end{figure}

Figure~\ref{choose_prob} shows that, 
by dynamically switching among different InterRL methods, our algorithm can achieve better performance though the whole learning process, fully demonstrating the benefits of interplay between different InterRL methods. In order to test the robustness, we added the Q-learning method into the portfolio of methods, and then manually disabled the functions of the other four individual InterRL methods in the 200th episode. Disabling the InterRL method means that the method provides a wrong guidance. Figure~\ref{worst:a} shows that by on-line monitoring the performance of each InterRL methods, the algorithm can detect any environment dynamics and adapt to this change in time.

Since individual InterRL methods tailor the human's guidance to agents with a particular representation and specialized shaping procedures, various parameter settings can bring about diverse learning performance when these methods are applied alone. In fact, none of the four individual InterRL methods can achieve a parameter-independent performance. Especially, human factors such as correctness and influence of human guidance play a crucial role in biasing the performance of each InterRL method. Generally, when the correctness of human guidance is not high enough, the individual InterRL methods are more likely to fail and may lead to divergence of learning process. Moreover, for the value-based and reward-based methods to be more efficient, human reward should be set to a relatively low value to reduce its influence. These rules provide basic principles to the selection of any individual methods in an InterRL setting. Our algorithm, however, due to the interplay between different methods, can take the benefits of each method to achieve a more robust and efficient learning performance.

\begin{figure}[t]
	\centering
	\subfigure[Average reward] {
		\includegraphics[width=0.22\textwidth]{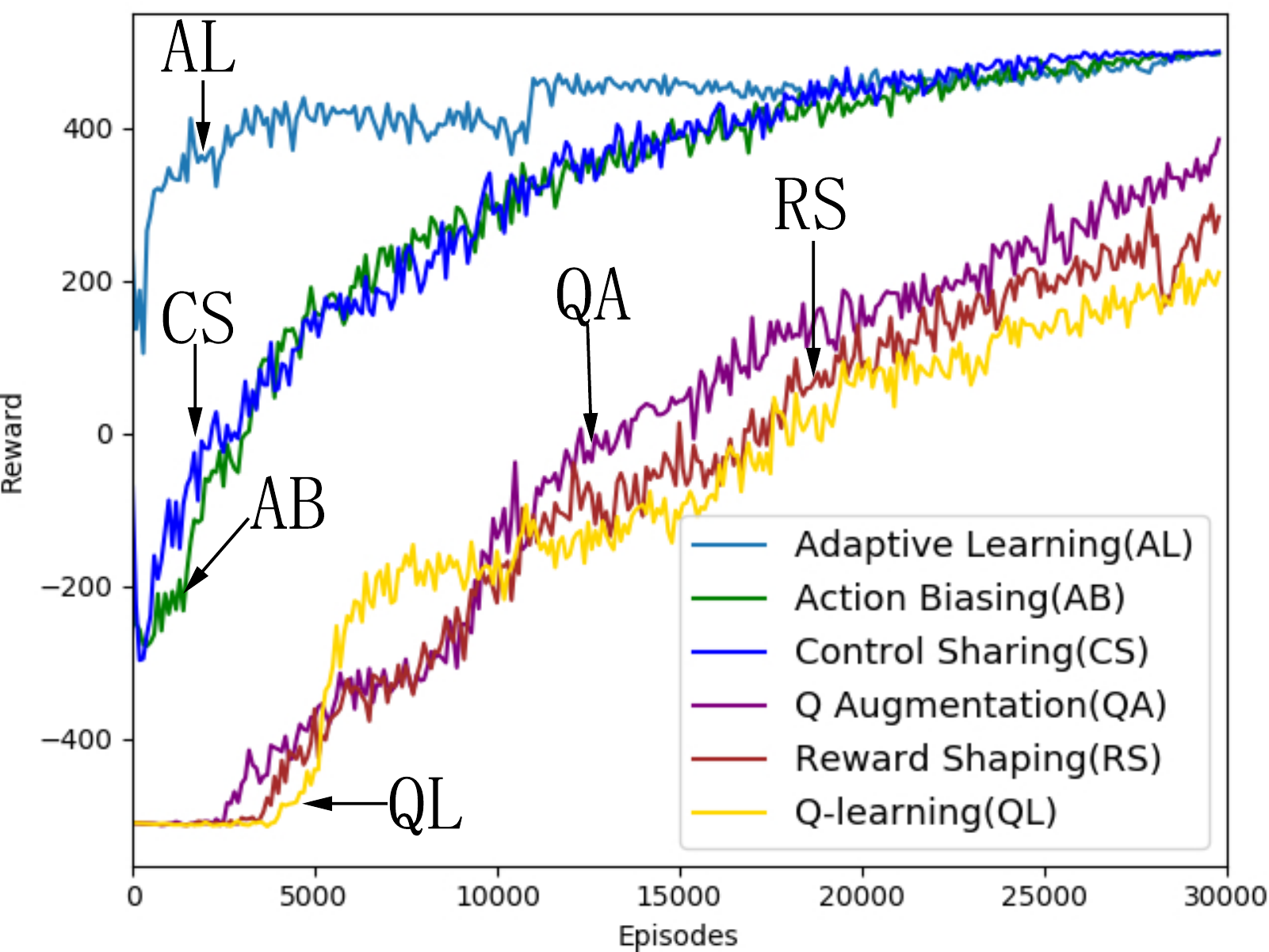}
		\label{fig1}
	}
	\subfigure[Interplay of methods] {
		\includegraphics[width=0.22\textwidth]{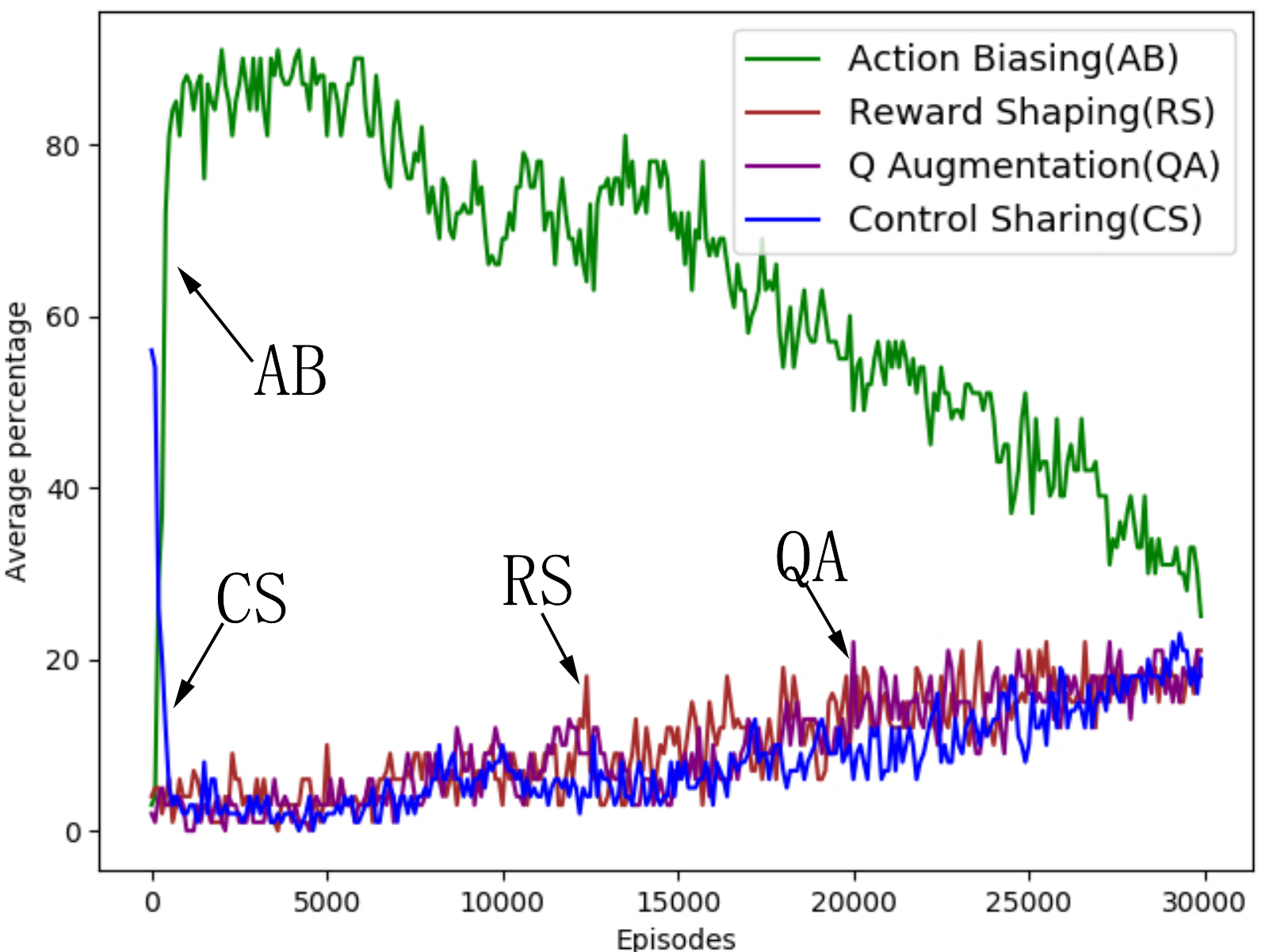}
	\label{fig2}
	}\\
	\subfigure[Average reward] {
		\includegraphics[width=0.22\textwidth]{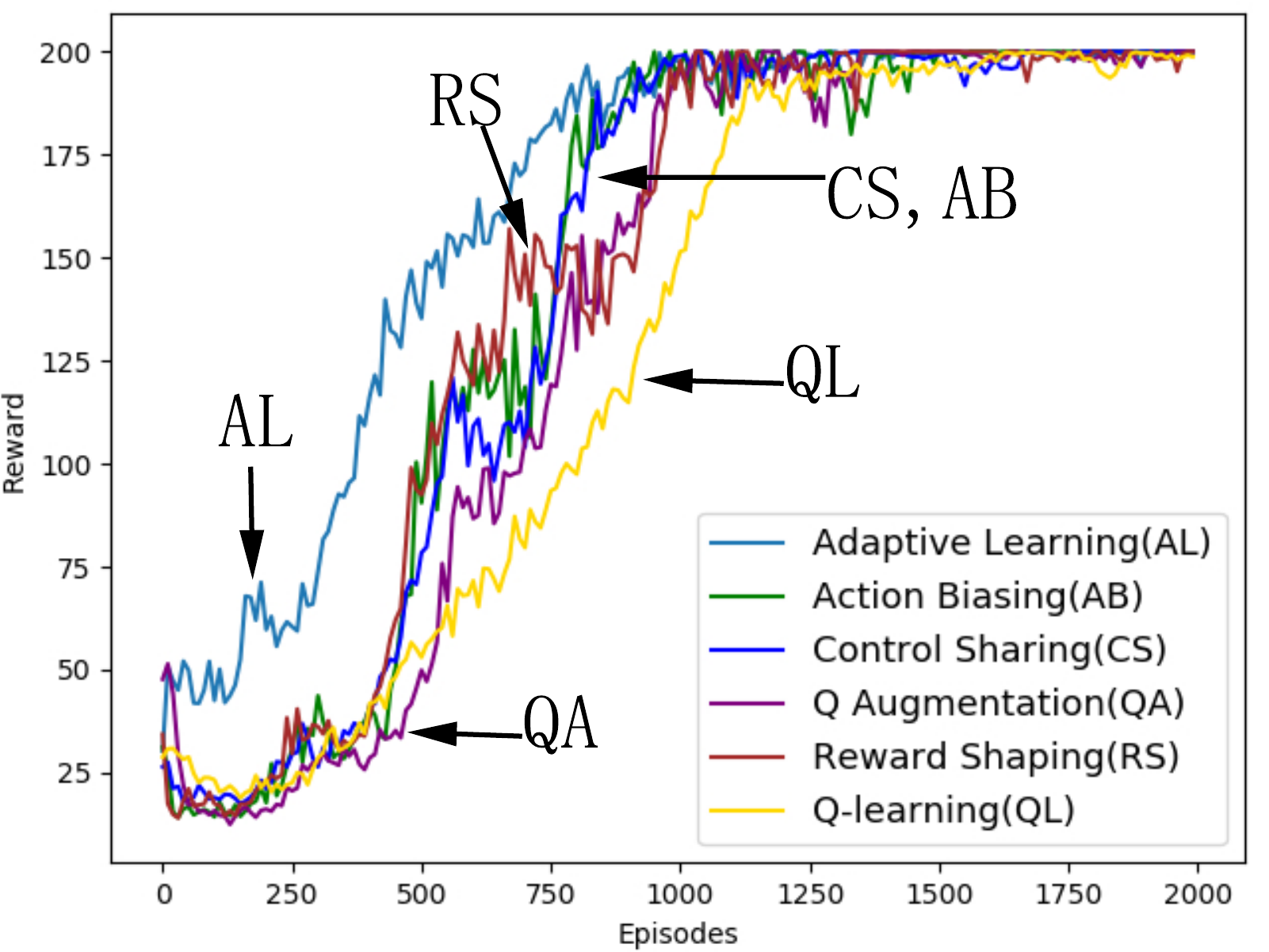}
		\label{fig3}
	}
	\subfigure[Interplay of methods] {
		\includegraphics[width=0.22\textwidth]{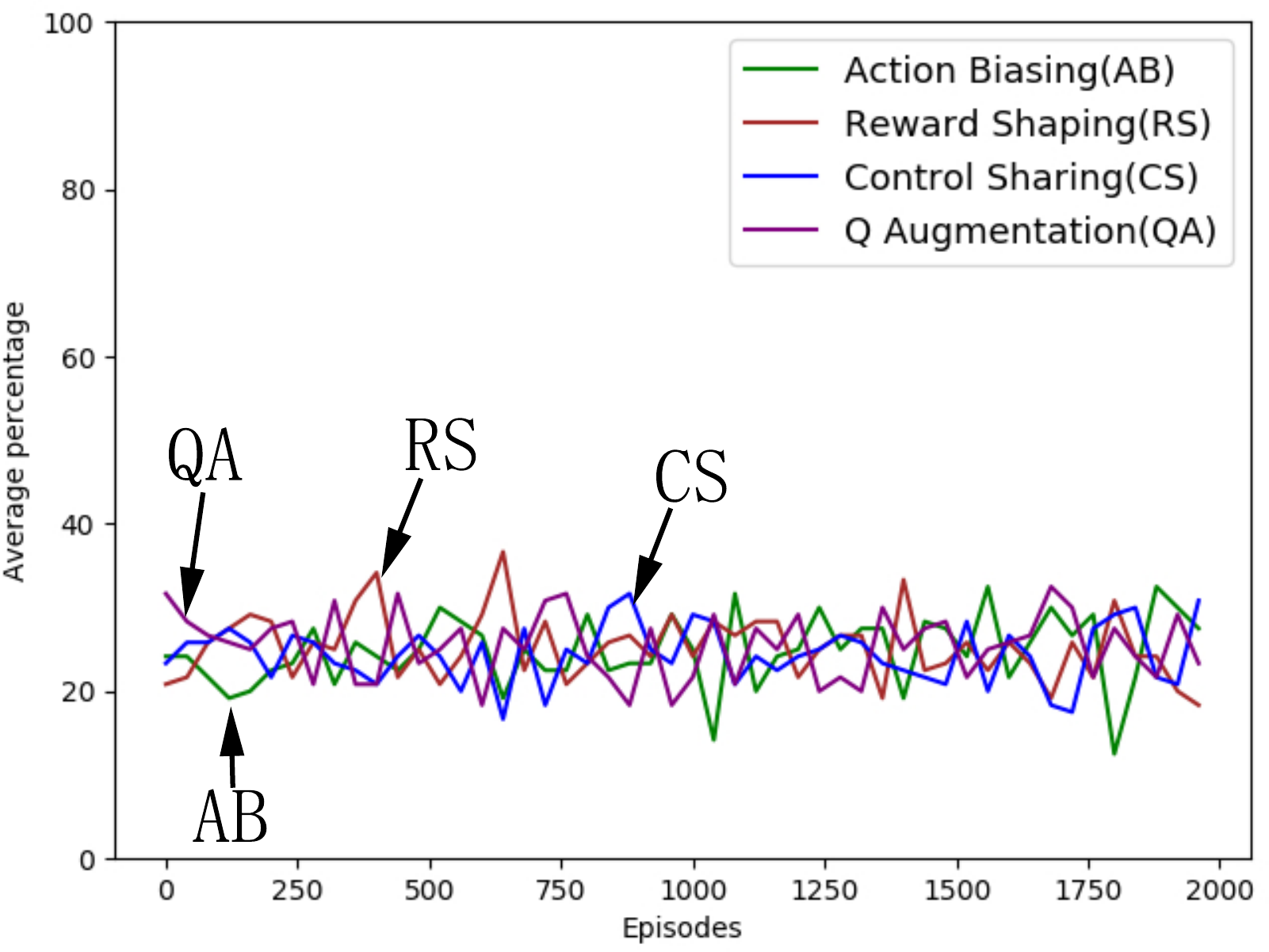}
		\label{fig4}
	}
	\caption{Results when real humans interact with RL agents in the Pac-Man domain (a)-(b), and the Cart-Pole domain (c)-(d). Each value in the Pac-Man and Cart-Pole domain is averaged over 100 and 10 episodes, respectively.}
	\label{real_humans}
\end{figure}

\subsection{Real Humans}
Figure~\ref{fig1} shows that the proposed adaptive learning method AL performs best among all the methods when real humans interact with the RL agent in the Pac-Man domain.
It is clear that methods directly shaping an RL agent's actions (AB) and policies (CS) are generally more efficient than those directly shaping the rewards (RS) and value functions (QA) in real human scenarios. Although the individual InterRL methods perform variously in the same parameter setting, the dynamic interplay of these methods can greatly promote learning performance. Figure~\ref{fig2} shows the average percentage of selecting the InterRL methods when using AL. Method AB is selected far more often than the other three methods throughout the learning process, but finally all the methods are chosen with an almost equal probability when the agent has converged. Figure~\ref{fig3} shows the performance in the Cart-Pole domain. Since continuous states in this domain can cause significant cognitive burdens and thus errors for humans, the individual methods can only achieve a slightly better performance against Q-learning, and an almost equal probability of selection as shown in Figure~\ref{fig4}.

The interplay of InterRL methods with the corresponding improved learning performance presents an interesting phenomenon. In the Pac-Man domain, although the \emph{action-based} method AB alone can already guarantee a good performance, it still faces the ``\emph{error curse}'' dilemma when an occasional error action given by the human can potentially bias the agent's learning process.
By slightly choosing other InterRL methods during the early stage, particularly those indirectly affecting the actions and policies (e.g., RS and QA), the ``\emph{error curse}'' dilemma can be greatly relieved, significantly promoting the overall learning performance using the AL method. Situations are a bit different in the Cart-Pole domain where the four individual methods perform similarly and only slightly better than Q-learning. Since the benefit of human learning is not as apparent as that in simpler discrete domains, no specific InterRL methods can dominate the dynamic interplay process.

\begin{figure}[t]
	\centering
	\subfigure[Oracle VS Real] {
		\includegraphics[width=0.22\textwidth]{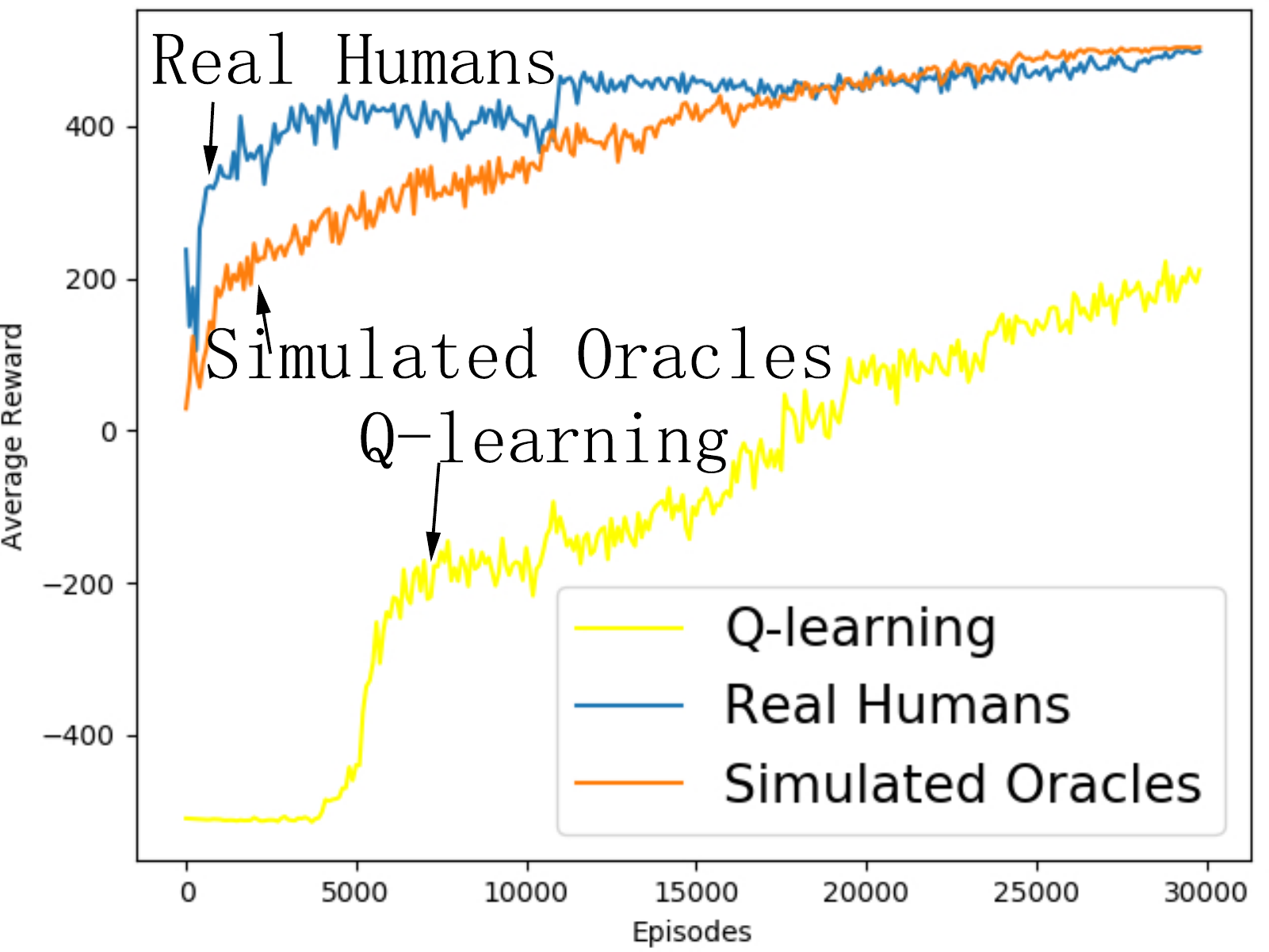}
		\label{fig5}
	}
	\subfigure[Training strategy] {
		\includegraphics[width=0.22\textwidth]{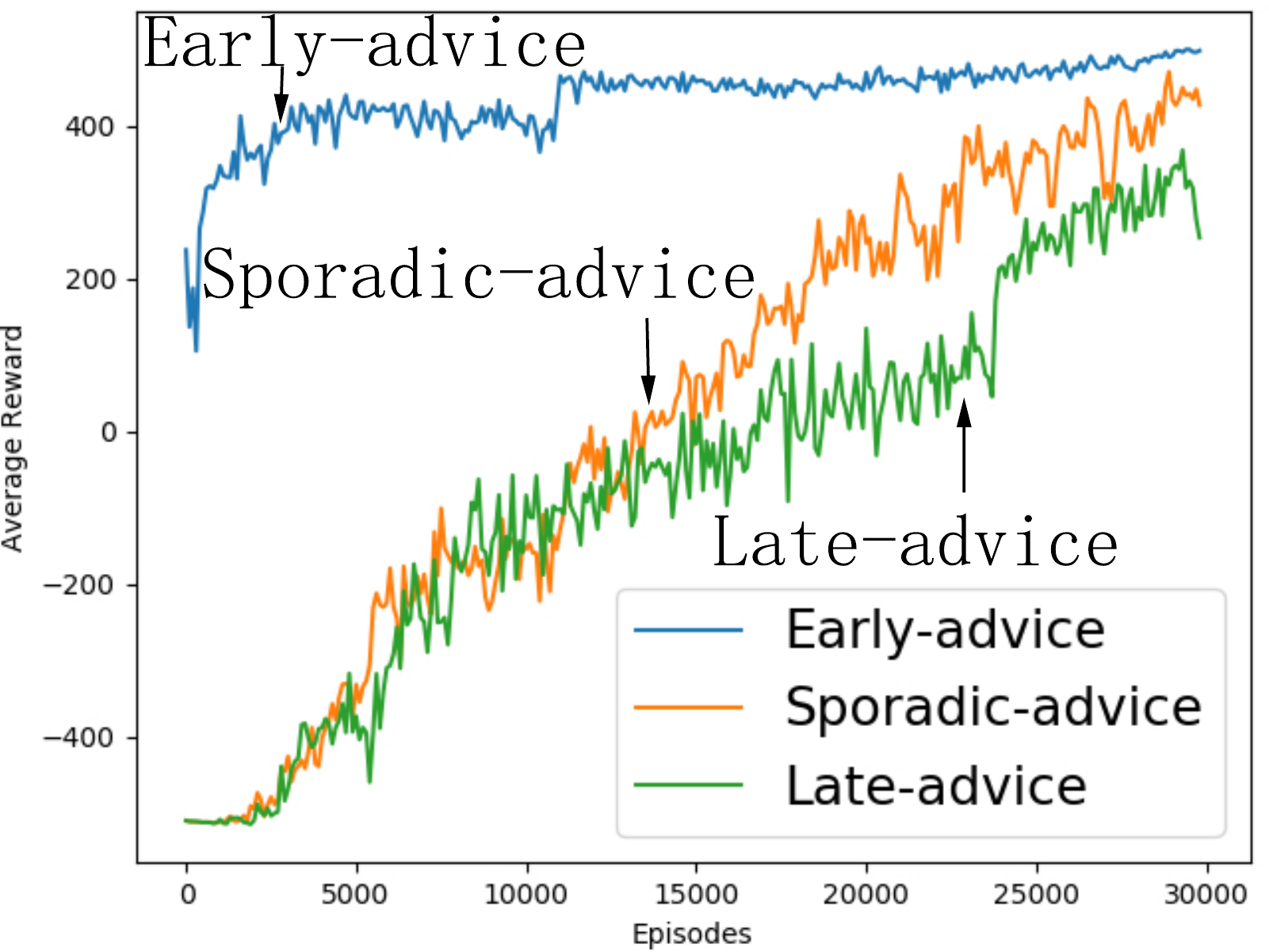}
		\label{fig6}
	}
	\caption{Performance of oracles against real humans and different training strategies in the Pac-Man domain. Each value is averaged over 100 episodes.}
	\label{real_humans}
\end{figure}

A more interesting yet a bit counterintuitive result is regarding the performance distinction between real humans and simulated oracles. Figure~\ref{fig5} shows that the simulated oracle produces lower quality performance than the real human when both methods use the same early-advice training strategy\footnote{To realize the same training strategy, we modify the $L$ parameter in simulated oracles to shape the agent in the first $L\times N$ episodes, where $N$ is the total number of training episodes.}. The result is a bit surprising since it is believed that the optimal strategy generated by the oracle should be more useful than the flawed data by the human. Upon a deeper investigation of the evaluations, we found that the simulated oracle gave clearly bad feedback in a number of states. For example, oracles seldom suggest moving towards the ghost, but humans might do this once in a while in order to get the food more quickly in the near future. This is because the oracle only knows one of the many optimal policies that wins, but does not recognize any other winning policies. If the action is different from what is recommended by its own winning policy, the simulated oracle simply gives bad critique on this action even though this action represents another way of winning the game. A human, however, might approve of several different good strategies at the same time, and can therefore beat the simulated oracle. Previous study has also observed such a phenomenon when humans and agents use the \emph{policy shaping} strategy \cite{cederborg2015policy}, suggesting the important role of human factors in an agent's RL process, that is, although human reinforcement is generally flawed, the informationally rich knowledge of human learning can bring about significant benefits over the flawless yet poor agent learning.


Figure~\ref{fig6} shows the performance when using different training strategies. It is clear that training at the early stage is far more efficient than training throughout the process or at later stage. The benefits of early advice has been supported by several previous studies \cite{knox2012reinforcement,torrey2013teaching}, although in different InterRL settings. To guarantee a maximum performance promotion, humans should step in the learning process as early as possible, before it is too late to entrench their influence in shaping the agent's learning behavior.

\section{Related Work}
There is plenty work on studying how humans can help in an agent's RL process \cite{taylor2018improving}. Various InterRL methods have been proposed. Knox \emph{et al.} \cite{knox2008tamer} proposed the TAMER framework for agents that can be interactively shaped by human trainers who give only positive and negative feedback signals. However, TAMER does not allow human rewards to be directly combined with autonomous learning reward. Later, Knox \emph{et al.} proposed eight plausible combination methods for combining a previously learned human reinforcement function with an MDP reward in RL \cite{knox2010combining}, and studied how human rewards and RL rewards can be combined simultaneously \cite{knox2012reinforcement}. Griffith \emph{et al.} \cite{griffith2013policy} introduced an algorithm for estimating a human's Bayes optimal policy and a technique for combining this with the policy formed from the agent's direct experience in the environment. Abel \emph{et al.} \cite{abel2017agent}) developed a general agent-agnostic framework for human-agent interaction that can capture a wide range of ways a human can help an RL agent. Other works focus on investigating when and how engagement of a human is valuable in InterRL settings \cite{li2013using,macglashan2017interactive,loftin2014strategy,mandel2017add}. All these methods, however, only target at specialized shaping procedures and their various learning performance in different settings. The most similar work to us is by Brys \emph{et al.} \cite{brys2017multi}, who applied ensembles of shaping methods to achieve multi-objectives in RL. However, only reward shaping methods were considered in their approach.

A large scale human study has been recently conducted by Rosenfeld et al. to investigate how human knowledge can help in tabular reinforcement learning \cite{rosenfeld2018leveraging}. However, in our design methodology, humans are coupled with the agent learning process at each time step during learning, which is a bit different from their work. Moreover, our work focuses on understanding the roles and advantages of different InterRL methods in shaping an agent's learning process.


There is also tremendous work that uses demonstrations or advice from humans or simulated agent teachers to facilitate RL. For example, the HAT algorithm transfers knowledge directly from human policies \cite{taylor2011integrating}. Other following work showed how expert advice or demonstrations can be used to shape rewards in an RL problem \cite{brys2015reinforcement,cederborg2015policy,chernova2016learning}. Some studies have also analyzed a teacher agent's action advise on RL under the teacher-student advising framework~\cite{fachantidis2017learning,torrey2013teaching}. However, all these studies focus more on transferring human/agent knowledge to RL, which differs slightly from our work that focuses on adaptively combining human rewards with RL rewards.

\section{Conclusions}

In human-in-the-loop InterRL, various human factors, such as different capabilities in solving the task, diversities in human knowledge and importance of human feedback can play a vital role in the final performance using different InterRL methods.
In this paper, we verified previous hypothesis that the interplay between different InterRL methods would potentially lead to new powerful shaping methods by taking advantage of the benefits of each InterRL method \cite{abel2017agent}. Results from both simulated and real human studies showed that the proposed adaptive shaping algorithm could guarantee high-level performance across a variety of domain settings. Future work includes more extensive evaluations on the adaptive shaping algorithm when real humans are characterized by various capabilities and influences.





\bibliographystyle{ACM-Reference-Format}  
\bibliography{arXiv_HumanReward}  


\begin{thebibliography}{00}


\ifx \showCODEN    \undefined \def \showCODEN     #1{\unskip}     \fi
\ifx \showDOI      \undefined \def \showDOI       #1{#1}\fi
\ifx \showISBNx    \undefined \def \showISBNx     #1{\unskip}     \fi
\ifx \showISBNxiii \undefined \def \showISBNxiii  #1{\unskip}     \fi
\ifx \showISSN     \undefined \def \showISSN      #1{\unskip}     \fi
\ifx \showLCCN     \undefined \def \showLCCN      #1{\unskip}     \fi
\ifx \shownote     \undefined \def \shownote      #1{#1}          \fi
\ifx \showarticletitle \undefined \def \showarticletitle #1{#1}   \fi
\ifx \showURL      \undefined \def \showURL       {\relax}        \fi
\providecommand\bibfield[2]{#2}
\providecommand\bibinfo[2]{#2}
\providecommand\natexlab[1]{#1}
\providecommand\showeprint[2][]{arXiv:#2}

\bibitem[\protect\citeauthoryear{Abel, Hershkowitz, Barth-Maron, Brawner,
  O'Farrell, MacGlashan, and Tellex}{Abel et~al\mbox{.}}{2015}]%
        {abel2015goal}
\bibfield{author}{\bibinfo{person}{David Abel}, \bibinfo{person}{D~Ellis
  Hershkowitz}, \bibinfo{person}{Gabriel Barth-Maron}, \bibinfo{person}{Stephen
  Brawner}, \bibinfo{person}{Kevin O'Farrell}, \bibinfo{person}{James
  MacGlashan}, {and} \bibinfo{person}{Stefanie Tellex}.}
  \bibinfo{year}{2015}\natexlab{}.
\newblock \showarticletitle{Goal-Based Action Priors.}. In
  \bibinfo{booktitle}{{\em ICAPS2015}}. \bibinfo{pages}{306--314}.
\newblock


\bibitem[\protect\citeauthoryear{Abel, Salvatier, Stuhlm{\"u}ller, and
  Evans}{Abel et~al\mbox{.}}{2017}]%
        {abel2017agent}
\bibfield{author}{\bibinfo{person}{David Abel}, \bibinfo{person}{John
  Salvatier}, \bibinfo{person}{Andreas Stuhlm{\"u}ller}, {and}
  \bibinfo{person}{Owain Evans}.} \bibinfo{year}{2017}\natexlab{}.
\newblock \showarticletitle{Agent-Agnostic Human-in-the-Loop Reinforcement
  Learning}.
\newblock \bibinfo{journal}{{\em arXiv preprint arXiv:1701.04079\/}}
  (\bibinfo{year}{2017}).
\newblock


\bibitem[\protect\citeauthoryear{Amir, Kamar, Kolobov, and Grosz}{Amir
  et~al\mbox{.}}{2016}]%
        {amir2016interactive}
\bibfield{author}{\bibinfo{person}{Ofra Amir}, \bibinfo{person}{Ece Kamar},
  \bibinfo{person}{Andrey Kolobov}, {and} \bibinfo{person}{Barbara Grosz}.}
  \bibinfo{year}{2016}\natexlab{}.
\newblock \showarticletitle{Interactive Teaching Strategies for Agent
  Training}. In \bibinfo{booktitle}{{\em IJCAI}}.
\newblock


\bibitem[\protect\citeauthoryear{Brys, Harutyunyan, Suay, Chernova, and
  Taylor}{Brys et~al\mbox{.}}{2015}]%
        {brys2015reinforcement}
\bibfield{author}{\bibinfo{person}{Tim Brys}, \bibinfo{person}{Anna
  Harutyunyan}, \bibinfo{person}{Halit~Bener Suay}, \bibinfo{person}{Sonia
  Chernova}, {and} \bibinfo{person}{Matthew~E. Taylor}.}
  \bibinfo{year}{2015}\natexlab{}.
\newblock \showarticletitle{Reinforcement learning from demonstration through
  shaping}. In \bibinfo{booktitle}{{\em IJCAI2015}}.
  \bibinfo{pages}{3352--3358}.
\newblock


\bibitem[\protect\citeauthoryear{Brys, Harutyunyan, Vrancx, Now{\'e}, and
  Taylor}{Brys et~al\mbox{.}}{2017}]%
        {brys2017multi}
\bibfield{author}{\bibinfo{person}{Tim Brys}, \bibinfo{person}{Anna
  Harutyunyan}, \bibinfo{person}{Peter Vrancx}, \bibinfo{person}{Ann Now{\'e}},
  {and} \bibinfo{person}{Matthew~E Taylor}.} \bibinfo{year}{2017}\natexlab{}.
\newblock \showarticletitle{Multi-objectivization and ensembles of shapings in
  reinforcement learning}.
\newblock \bibinfo{journal}{{\em Neurocomputing\/}}  \bibinfo{volume}{263}
  (\bibinfo{year}{2017}), \bibinfo{pages}{48--59}.
\newblock


\bibitem[\protect\citeauthoryear{Cederborg, Grover, Isbell, and
  Thomaz}{Cederborg et~al\mbox{.}}{2015}]%
        {cederborg2015policy}
\bibfield{author}{\bibinfo{person}{Thomas Cederborg}, \bibinfo{person}{Ishaan
  Grover}, \bibinfo{person}{Charles~L Isbell}, {and}
  \bibinfo{person}{Andrea~Lockerd Thomaz}.} \bibinfo{year}{2015}\natexlab{}.
\newblock \showarticletitle{Policy Shaping with Human Teachers.}. In
  \bibinfo{booktitle}{{\em IJCAI}}. \bibinfo{pages}{3366--3372}.
\newblock


\bibitem[\protect\citeauthoryear{Chernova, Chernova, Chernova, and
  Chernova}{Chernova et~al\mbox{.}}{2016}]%
        {chernova2016learning}
\bibfield{author}{\bibinfo{person}{Sonia Chernova}, \bibinfo{person}{Sonia
  Chernova}, \bibinfo{person}{Sonia Chernova}, {and} \bibinfo{person}{Sonia
  Chernova}.} \bibinfo{year}{2016}\natexlab{}.
\newblock \showarticletitle{Learning from Demonstration for Shaping through
  Inverse Reinforcement Learning}. In \bibinfo{booktitle}{{\em AAMAS2016}}.
  \bibinfo{pages}{429--437}.
\newblock


\bibitem[\protect\citeauthoryear{Devlin and Kudenko}{Devlin and
  Kudenko}{2012}]%
        {devlin2012dynamic}
\bibfield{author}{\bibinfo{person}{Sam Devlin} {and} \bibinfo{person}{Daniel
  Kudenko}.} \bibinfo{year}{2012}\natexlab{}.
\newblock \showarticletitle{Dynamic potential-based reward shaping}. In
  \bibinfo{booktitle}{{\em AAMAS2012}}. \bibinfo{pages}{433--440}.
\newblock


\bibitem[\protect\citeauthoryear{Fachantidis, Taylor, and Vlahavas}{Fachantidis
  et~al\mbox{.}}{2017}]%
        {fachantidis2017learning}
\bibfield{author}{\bibinfo{person}{Anestis Fachantidis},
  \bibinfo{person}{Matthew~E Taylor}, {and} \bibinfo{person}{Ioannis
  Vlahavas}.} \bibinfo{year}{2017}\natexlab{}.
\newblock \showarticletitle{Learning to Teach Reinforcement Learning Agents}.
\newblock \bibinfo{journal}{{\em Machine Learning and Knowledge Extraction\/}}
  \bibinfo{volume}{1}, \bibinfo{number}{1} (\bibinfo{year}{2017}),
  \bibinfo{pages}{2}.
\newblock


\bibitem[\protect\citeauthoryear{Fern{\'a}ndez and Veloso}{Fern{\'a}ndez and
  Veloso}{2006}]%
        {fernandez2006probabilistic}
\bibfield{author}{\bibinfo{person}{Fernando Fern{\'a}ndez} {and}
  \bibinfo{person}{Manuela Veloso}.} \bibinfo{year}{2006}\natexlab{}.
\newblock \showarticletitle{Probabilistic policy reuse in a reinforcement
  learning agent}. In \bibinfo{booktitle}{{\em AAMAS2006}}. ACM,
  \bibinfo{pages}{720--727}.
\newblock


\bibitem[\protect\citeauthoryear{Griffith, Subramanian, Scholz, Isbell, and
  Thomaz}{Griffith et~al\mbox{.}}{2013}]%
        {griffith2013policy}
\bibfield{author}{\bibinfo{person}{Shane Griffith}, \bibinfo{person}{Kaushik
  Subramanian}, \bibinfo{person}{Jonathan Scholz}, \bibinfo{person}{Charles~L
  Isbell}, {and} \bibinfo{person}{Andrea~L Thomaz}.}
  \bibinfo{year}{2013}\natexlab{}.
\newblock \showarticletitle{Policy shaping: Integrating human feedback with
  reinforcement learning}. In \bibinfo{booktitle}{{\em NIPS2013}}.
  \bibinfo{pages}{2625--2633}.
\newblock


\bibitem[\protect\citeauthoryear{Jennings, Moreau, Nicholson, Ramchurn,
  Roberts, Rodden, and Rogers}{Jennings et~al\mbox{.}}{2014}]%
        {jennings2014human}
\bibfield{author}{\bibinfo{person}{Nicholas~R Jennings}, \bibinfo{person}{Luc
  Moreau}, \bibinfo{person}{David Nicholson}, \bibinfo{person}{Sarvapali
  Ramchurn}, \bibinfo{person}{Stephen Roberts}, \bibinfo{person}{Tom Rodden},
  {and} \bibinfo{person}{Alex Rogers}.} \bibinfo{year}{2014}\natexlab{}.
\newblock \showarticletitle{Human-agent collectives}.
\newblock \bibinfo{journal}{{\it Commun. ACM}} \bibinfo{volume}{57},
  \bibinfo{number}{12} (\bibinfo{year}{2014}), \bibinfo{pages}{80--88}.
\newblock


\bibitem[\protect\citeauthoryear{Judah, Roy, Fern, and Dietterich}{Judah
  et~al\mbox{.}}{2010}]%
        {judah2010reinforcement}
\bibfield{author}{\bibinfo{person}{Kshitij Judah}, \bibinfo{person}{Saikat
  Roy}, \bibinfo{person}{Alan Fern}, {and} \bibinfo{person}{Thomas~G
  Dietterich}.} \bibinfo{year}{2010}\natexlab{}.
\newblock \showarticletitle{Reinforcement Learning Via Practice and Critique
  Advice.}. In \bibinfo{booktitle}{{\em AAAI}}.
\newblock


\bibitem[\protect\citeauthoryear{Knox and Stone}{Knox and Stone}{2008}]%
        {knox2008tamer}
\bibfield{author}{\bibinfo{person}{W~Bradley Knox} {and} \bibinfo{person}{Peter
  Stone}.} \bibinfo{year}{2008}\natexlab{}.
\newblock \showarticletitle{Tamer: Training an agent manually via evaluative
  reinforcement}. In \bibinfo{booktitle}{{\em Development and Learning, 2008.
  ICDL 2008. 7th IEEE International Conference on}}. IEEE,
  \bibinfo{pages}{292--297}.
\newblock


\bibitem[\protect\citeauthoryear{Knox and Stone}{Knox and Stone}{2010}]%
        {knox2010combining}
\bibfield{author}{\bibinfo{person}{W~Bradley Knox} {and} \bibinfo{person}{Peter
  Stone}.} \bibinfo{year}{2010}\natexlab{}.
\newblock \showarticletitle{Combining manual feedback with subsequent MDP
  reward signals for reinforcement learning}. In \bibinfo{booktitle}{{\em
  AAMAS2010}}. \bibinfo{pages}{5--12}.
\newblock


\bibitem[\protect\citeauthoryear{Knox and Stone}{Knox and Stone}{2012}]%
        {knox2012reinforcement}
\bibfield{author}{\bibinfo{person}{W~Bradley Knox} {and} \bibinfo{person}{Peter
  Stone}.} \bibinfo{year}{2012}\natexlab{}.
\newblock \showarticletitle{Reinforcement learning from simultaneous human and
  MDP reward}. In \bibinfo{booktitle}{{\em AAMAS2012}}.
  \bibinfo{pages}{475--482}.
\newblock


\bibitem[\protect\citeauthoryear{Knox and Stone}{Knox and Stone}{2015}]%
        {knox2015framing}
\bibfield{author}{\bibinfo{person}{W~Bradley Knox} {and} \bibinfo{person}{Peter
  Stone}.} \bibinfo{year}{2015}\natexlab{}.
\newblock \showarticletitle{Framing reinforcement learning from human reward:
  Reward positivity, temporal discounting, episodicity, and performance}.
\newblock \bibinfo{journal}{{\em Artificial Intelligence\/}}
  \bibinfo{volume}{225} (\bibinfo{year}{2015}), \bibinfo{pages}{24--50}.
\newblock


\bibitem[\protect\citeauthoryear{Li, Hung, Whiteson, and Knox}{Li
  et~al\mbox{.}}{2013}]%
        {li2013using}
\bibfield{author}{\bibinfo{person}{Guangliang Li}, \bibinfo{person}{Hayley
  Hung}, \bibinfo{person}{Shimon Whiteson}, {and} \bibinfo{person}{W~Bradley
  Knox}.} \bibinfo{year}{2013}\natexlab{}.
\newblock \showarticletitle{Using informative behavior to increase engagement
  in the tamer framework}. In \bibinfo{booktitle}{{\em AAMAS2013}}.
  \bibinfo{pages}{909--916}.
\newblock


\bibitem[\protect\citeauthoryear{Loftin, MacGlashan, Peng, Taylor, Littman,
  Huang, and Roberts}{Loftin et~al\mbox{.}}{2014}]%
        {loftin2014strategy}
\bibfield{author}{\bibinfo{person}{Robert~Tyler Loftin}, \bibinfo{person}{James
  MacGlashan}, \bibinfo{person}{Bei Peng}, \bibinfo{person}{Matthew~E Taylor},
  \bibinfo{person}{Michael~L Littman}, \bibinfo{person}{Jeff Huang}, {and}
  \bibinfo{person}{David~L Roberts}.} \bibinfo{year}{2014}\natexlab{}.
\newblock \showarticletitle{A Strategy-Aware Technique for Learning Behaviors
  from Discrete Human Feedback.}. In \bibinfo{booktitle}{{\em AAAI}}.
  \bibinfo{pages}{937--943}.
\newblock


\bibitem[\protect\citeauthoryear{MacGlashan, Ho, Loftin, Peng, Roberts, Taylor,
  and Littman}{MacGlashan et~al\mbox{.}}{2017}]%
        {macglashan2017interactive}
\bibfield{author}{\bibinfo{person}{James MacGlashan}, \bibinfo{person}{Mark~K
  Ho}, \bibinfo{person}{Robert Loftin}, \bibinfo{person}{Bei Peng},
  \bibinfo{person}{David Roberts}, \bibinfo{person}{Matthew~E Taylor}, {and}
  \bibinfo{person}{Michael~L Littman}.} \bibinfo{year}{2017}\natexlab{}.
\newblock \showarticletitle{Interactive learning from policy-dependent human
  feedback}.
\newblock \bibinfo{journal}{{\em arXiv preprint arXiv:1701.06049\/}}
  (\bibinfo{year}{2017}).
\newblock


\bibitem[\protect\citeauthoryear{Mandel, Liu, Brunskill, and Popovic}{Mandel
  et~al\mbox{.}}{2017}]%
        {mandel2017add}
\bibfield{author}{\bibinfo{person}{Travis Mandel}, \bibinfo{person}{Yun-En
  Liu}, \bibinfo{person}{Emma Brunskill}, {and} \bibinfo{person}{Zoran
  Popovic}.} \bibinfo{year}{2017}\natexlab{}.
\newblock \showarticletitle{Where to Add Actions in Human-in-the-Loop
  Reinforcement Learning.}. In \bibinfo{booktitle}{{\em AAAI}}.
  \bibinfo{pages}{2322--2328}.
\newblock


\bibitem[\protect\citeauthoryear{Peng, MacGlashan, Loftin, Littman, Roberts,
  and Taylor}{Peng et~al\mbox{.}}{2016}]%
        {peng2016need}
\bibfield{author}{\bibinfo{person}{Bei Peng}, \bibinfo{person}{James
  MacGlashan}, \bibinfo{person}{Robert Loftin}, \bibinfo{person}{Michael~L
  Littman}, \bibinfo{person}{David~L Roberts}, {and} \bibinfo{person}{Matthew~E
  Taylor}.} \bibinfo{year}{2016}\natexlab{}.
\newblock \showarticletitle{A need for speed: Adapting agent action speed to
  improve task learning from non-expert humans}. In \bibinfo{booktitle}{{\em
  AAMAS2016}}. \bibinfo{pages}{957--965}.
\newblock


\bibitem[\protect\citeauthoryear{Rosenfeld, Cohen, Taylor, and Kraus}{Rosenfeld
  et~al\mbox{.}}{2018}]%
        {rosenfeld2018leveraging}
\bibfield{author}{\bibinfo{person}{Ariel Rosenfeld}, \bibinfo{person}{Moshe
  Cohen}, \bibinfo{person}{Matthew~E Taylor}, {and} \bibinfo{person}{Sarit
  Kraus}.} \bibinfo{year}{2018}\natexlab{}.
\newblock \showarticletitle{Leveraging human knowledge in tabular reinforcement
  learning: A study of human subjects}.
\newblock \bibinfo{journal}{{\em arXiv preprint arXiv:1805.05769\/}}
  (\bibinfo{year}{2018}).
\newblock


\bibitem[\protect\citeauthoryear{Sherstov and Stone}{Sherstov and
  Stone}{2005}]%
        {sherstov2005improving}
\bibfield{author}{\bibinfo{person}{Alexander~A Sherstov} {and}
  \bibinfo{person}{Peter Stone}.} \bibinfo{year}{2005}\natexlab{}.
\newblock \showarticletitle{Improving action selection in MDP's via knowledge
  transfer}. In \bibinfo{booktitle}{{\em AAAI2005}}, Vol.~\bibinfo{volume}{5}.
  \bibinfo{pages}{1024--1029}.
\newblock


\bibitem[\protect\citeauthoryear{Sutton and Barto}{Sutton and Barto}{1998}]%
        {sutton1998reinforcement}
\bibfield{author}{\bibinfo{person}{R.S. Sutton} {and} \bibinfo{person}{A.G.
  Barto}.} \bibinfo{year}{1998}\natexlab{}.
\newblock \bibinfo{booktitle}{{\em {Reinforcement learning: An introduction}}}.
\newblock \bibinfo{publisher}{The MIT press}.
\newblock
\showISBNx{0262193981}


\bibitem[\protect\citeauthoryear{Taylor and Borealis}{Taylor and
  Borealis}{2018}]%
        {taylor2018improving}
\bibfield{author}{\bibinfo{person}{Matthew~E Taylor} {and} \bibinfo{person}{AI
  Borealis}.} \bibinfo{year}{2018}\natexlab{}.
\newblock \showarticletitle{Improving Reinforcement Learning with Human Input}.
  In \bibinfo{booktitle}{{\em IJCAI}}. \bibinfo{pages}{5724--5728}.
\newblock


\bibitem[\protect\citeauthoryear{Taylor, Suay, and Chernova}{Taylor
  et~al\mbox{.}}{2011}]%
        {taylor2011integrating}
\bibfield{author}{\bibinfo{person}{Matthew~E Taylor},
  \bibinfo{person}{Halit~Bener Suay}, {and} \bibinfo{person}{Sonia Chernova}.}
  \bibinfo{year}{2011}\natexlab{}.
\newblock \showarticletitle{Integrating reinforcement learning with human
  demonstrations of varying ability}. In \bibinfo{booktitle}{{\em AAMAS2011}}.
  \bibinfo{pages}{617--624}.
\newblock


\bibitem[\protect\citeauthoryear{Thomaz and Breazeal}{Thomaz and
  Breazeal}{2008}]%
        {thomaz2008teachable}
\bibfield{author}{\bibinfo{person}{Andrea~L Thomaz} {and}
  \bibinfo{person}{Cynthia Breazeal}.} \bibinfo{year}{2008}\natexlab{}.
\newblock \showarticletitle{Teachable robots: Understanding human teaching
  behavior to build more effective robot learners}.
\newblock \bibinfo{journal}{{\em Artificial Intelligence\/}}
  \bibinfo{volume}{172}, \bibinfo{number}{6-7} (\bibinfo{year}{2008}),
  \bibinfo{pages}{716--737}.
\newblock


\bibitem[\protect\citeauthoryear{Thomaz, Breazeal, et~al\mbox{.}}{Thomaz
  et~al\mbox{.}}{2006}]%
        {thomaz2006reinforcement}
\bibfield{author}{\bibinfo{person}{Andrea~Lockerd Thomaz},
  \bibinfo{person}{Cynthia Breazeal}, {et~al\mbox{.}}}
  \bibinfo{year}{2006}\natexlab{}.
\newblock \showarticletitle{Reinforcement learning with human teachers:
  Evidence of feedback and guidance with implications for learning
  performance}. In \bibinfo{booktitle}{{\em AAAI}}, Vol.~\bibinfo{volume}{6}.
  Boston, MA, \bibinfo{pages}{1000--1005}.
\newblock


\bibitem[\protect\citeauthoryear{Thomaz, Hoffman, and Breazeal}{Thomaz
  et~al\mbox{.}}{2005}]%
        {thomaz2005real}
\bibfield{author}{\bibinfo{person}{Andrea~Lockerd Thomaz}, \bibinfo{person}{Guy
  Hoffman}, {and} \bibinfo{person}{Cynthia Breazeal}.}
  \bibinfo{year}{2005}\natexlab{}.
\newblock \showarticletitle{Real-time interactive reinforcement learning for
  robots}. In \bibinfo{booktitle}{{\em AAAI 2005 workshop on human
  comprehensible machine learning}}.
\newblock


\bibitem[\protect\citeauthoryear{Torrey and Taylor}{Torrey and Taylor}{2013}]%
        {torrey2013teaching}
\bibfield{author}{\bibinfo{person}{Lisa Torrey} {and} \bibinfo{person}{Matthew
  Taylor}.} \bibinfo{year}{2013}\natexlab{}.
\newblock \showarticletitle{Teaching on a budget: Agents advising agents in
  reinforcement learning}. In \bibinfo{booktitle}{{\em AAMAS2013}}.
  \bibinfo{pages}{1053--1060}.
\newblock


\bibitem[\protect\citeauthoryear{Vien and Ertel}{Vien and Ertel}{2012}]%
        {vien2012reinforcement}
\bibfield{author}{\bibinfo{person}{Ngo~Anh Vien} {and}
  \bibinfo{person}{Wolfgang Ertel}.} \bibinfo{year}{2012}\natexlab{}.
\newblock \showarticletitle{Reinforcement learning combined with human feedback
  in continuous state and action spaces}. In \bibinfo{booktitle}{{\em
  Development and Learning and Epigenetic Robotics (ICDL), 2012 IEEE
  International Conference on}}. \bibinfo{pages}{1--6}.
\newblock


\end{thebibliography}

\end{document}